\documentclass[aps,prb,twocolumn,groupedaddress]{revtex4-2}
\usepackage{amsmath}
\usepackage{graphicx}
\usepackage{dcolumn}
\usepackage{bbm}
\usepackage{subfigure}
\usepackage{amssymb}
\usepackage{textcomp, gensymb}
\usepackage{latexsym}
\usepackage{makecell}
\usepackage{footnote}
\usepackage{soul}
\usepackage{bm}

\usepackage{color}
\usepackage[colorlinks,urlcolor=blue,linkcolor=blue,anchorcolor=blue,citecolor=blue]{hyperref}

\newcommand{\Sv}{\vec{S}}
\newcommand{\Sh}{{\hat S}}
\newcommand{\nh}{{\hat n}}
\newcommand{\mh}{{\hat m}}
\newcommand{\lh}{{\hat \ell}}
\newcommand{\eh}{{\hat e}}
\newcommand{\psih}{{\hat\psi}}
\newcommand{\cH}{{\mathcal H}}
\newcommand{\oh}{{\frac{1}{2}}}
\newcommand{\qv}{{\bf q}}
\newcommand{\xv}{{\bf x}}

\begin{document}


\title{\texorpdfstring{$O(N)$}{O(N)} smectic \texorpdfstring{$\sigma$}{sigma}-model}

\author{Tzu-Chi~Hsieh}
\author{Leo~Radzihovsky}

\affiliation{Department of Physics and Center
  for Theory of Quantum Matter, University of Colorado, Boulder, Colorado 80309, USA}

\date{\today}


\begin{abstract}
  A unidirectional ``density'' wave order in an otherwise {\em
    isotropic} environment is guaranteed to display smectic-like
  Goldstone mode. Examples of such “soft” states include conventional
  smectic liquid crystals, putative Fulde-Ferrell-Larkin-Ovchinnikov
  superfluids, and helical states of frustrated bosons and spins. Here we develop
  generalized spin-smectic $\sigma$-models that
  break $O(N)$ internal symmetry in addition to the $d$-dimensional
  rotational and uniaxial translational symmetries. We explore
  long-wavelength properties of such strongly fluctuating states, show
  that they are characterized by a ``double-power-law'' static
  structure peak, and analyze their asymptotic symmetry-reduced
  crossover to conventional low-energy modes.  We also present the
  associated Ginzburg-Landau theory, describing phase transition into
  such spin-smectic states, and discuss experimental realization of
  such models.
\end{abstract}

\maketitle


\section{Introduction}

\subsection{Background and motivation}
There are number of systems in nature that spontaneously undergo
transitions to a variety of ``density'' \footnote{Throughout the
  manuscript we will use the term ``density'' to also refer to spin,
  Cooper pair, and other generalized densities that transform
  nontrivially under internal symmetry.} wave states, characterized by
a periodic modulation. Such systems range from a large variety of
charge and spin density waves, e.g., antiferromagnetic insulators in
cuprates \cite{chakravartyTwodimensionalQuantumHeisenberg1989}, charge
density waves in $\text{NbSe}_\text{3}$
\cite{leeElectricFieldDepinning1979} and Wigner crystal
\cite{wignerInteractionElectronsMetals1934}, to the putative
pair-density wave superconductors
\cite{agterbergPhysicsPairDensityWaves2020}. Generically, in an
orientationally-ordered environment of a crystal, the low-energy
nonlinear $\sigma$-model (n$\sigma$m) description of the corresponding
Goldstone modes is given by their linear gradient elasticity, at weak
coupling controlled by a Gaussian fixed point, and is thus quite
well-understood
\cite{polyakovInteractionGoldstoneParticles1975,nelsonMomentumshellRecursionRelations1977,haldaneNonlinearFieldTheory1983,haldaneNonlinearEnsuremathSigma1988,chakravartyTwodimensionalQuantumHeisenberg1989,dombreNonlinearModelsTriangular1989}.

In contrast to these systems, there exists a qualitatively distinct
class of phases of matter, where a unidirectional density-wave order
(characterized by a spontaneously-oriented wavevector ${\bf q}_0$)
takes place in an isotropic environment, and thus, in addition to
translational and internal symmetries, also spontaneously breaks {\em
  spatial rotational} symmetry. We expect such states to be described
by a ``soft'' -- higher gradient -- n$\sigma$m that is qualitatively
distinct from their crystalline counterparts. The derivation and study
of such models are the focus of the present work.

The simplest realization of such states is the extensively-studied
conventional smectic liquid crystals \cite{gennesPhysicsLiquidCrystals1995}
characterized by a periodic {\em scalar} number-density, described by
a fully rotationally-invariant soft nonlinear elastic model
\cite{grinsteinAnharmonicEffectsBulk1981,grinsteinNonlinearElasticTheory1982b}. In
the presence of thermal and quantum fluctuations or quenched
disorder \cite{radzihovskyNematicSmecticMathit1997,radzihovskySmecticLiquidCrystals1999,belliniUniversalityScalingDisordering2001},
the corresponding scalar phonon (denoted as $u$) fluctuations are qualitatively
enhanced (e.g., for spatial dimension $d\le 3$, thermal $u_{rms}$
grows with system size, diverging in the thermodynamic limit) and
lead to importance of the nonlinear
elasticity \cite{grinsteinAnharmonicEffectsBulk1981}, resulting in a
tuning-free critical smectic phase described by universal
exponents \cite{2023arXiv230603142R}.

In this paper, we consider a generalization of such a scalar smectic
to ``spin'' (non-scalar) density, $\vec{S}({\bf x})$, that transforms
nontrivially under $O(N)$-spin rotational symmetry. Such a state, that
we dub ``spin-smectic", is characterized by an internal flavor degrees
of freedom, and exhibits a spontaneous uniaxial spatial modulation in
$\vec{S}({\bf x})$ along ${\bf q}_0$. It thereby spontaneously breaks
the $O(N)$-internal spin symmetry, and the underlying spatial
$O(d)$-rotational and $T(d)$-translational symmetries.

Our study of such spin-smectics is motivated by a number of physical
realizations of unidirectional orders, that include: (i) $O(N=1)$:
conventional smectic liquid crystal
\cite{gennesPhysicsLiquidCrystals1995},
and quantum Hall striped states of a two-dimensional electron gas at
half-filled high Landau levels
\cite{lillyEvidenceAnisotropicState1999,koulakovChargeDensityWave1996,moessnerExactResultsInteracting1996,fradkinLiquidcrystalPhasesQuantum1999,macdonaldQuantumTheoryQuantum2000,radzihovskyTheoryQuantumHall2002},
(ii) $O(N=2)$: cholesteric liquid crystal \cite{radzihovskyNonlinearSmecticElasticity2011}, spin-orbit coupled and dipolar Bose condensates
\cite{zhaiDegenerateQuantumGases2015a,
  tanziSupersolidSymmetryBreaking2019}, putative
Fulde-Ferrell-Larkin-Ovchinnikov (FFLO) paired superfluids
\cite{fuldeSuperconductivityStrongSpinExchange1964,osti_4653415}
in imbalanced degenerate atomic gases
\cite{radzihovskyQuantumLiquidCrystals2009,radzihovskyFluctuationsPhaseTransitions2011a}, a
$p$-wave resonant Bose gas
\cite{radzihovskyWaveResonantBose2009,choiFinitemomentumSuperfluidityPhase2011a},
and helical states of frustrated bosons
\cite{hsiehHelicalSuperfluidFrustrated2022}, (iii) $O(N=3)$:
helimagnets of frustrated spin systems \footnote{These are distinct
  from another interesting class of helimagnets arising from the
  chiral Dzyaloshinskii-Moriya (DM) spin-orbit interaction that drives
  the helical structure. Such DM helical states are less symmetrical than
those  considered in this paper, and do {\em not} exhibit independent
  $O(N)$ and $O(d)$ symmetries. In such systems, equivalent to a
  cholesteric liquid crystal the orientations of ${\bf q}_0$ and
  $\vec{S}$ are locked, and thus at low energies its Goldstone mode
  reduces to that of a smectic, as discussed in
  Ref.~\cite{radzihovskyNonlinearSmecticElasticity2011}. Physical
  realizations of such system include $\text{MnSi}$, which is an
  itinerant ferromagnet exhibiting helical structure at low
  temperatures. These systems show anomalous magnetic
  \cite{pfleidererPartialOrderNonFermiliquid2004,grigorievCriticalFluctuationsMnSi2005a,muhlbauerSkyrmionLatticeChiral2009}
  and transport \cite{neubauerTopologicalHallEffect2009} behavior as
  probed over different temperature, pressure, and magnetic field.}
\cite{bergmanOrderbydisorderSpiralSpinliquid2007,mulderSpiralOrderDisorder2010},
as realized in spinel materials, e.g.,
$\text{CoAl}_\text{2}\text{O}_\text{4}$
\cite{tristanGeometricFrustrationCubic2005,suzukiMeltingAntiferromagneticOrdering2007}
and $\text{MnSc}_\text{2}\text{S}_\text{4}$
\cite{fritschSpinOrbitalFrustration2004}, van der Waals honeycomb
magnets, e.g., $\text{FeCl}_\text{3}$ \cite{gaoSpiralSpinLiquid2022},
and stretched diamond lattice, e.g., $\text{LiYbO}_\text{2}$
\cite{bordelonFrustratedHeisenbergEnsuremath2021}. For long wavelength
$2\pi/q_0\gg a$ (lattice constant) spin-density modulation, even in the crystalline realizations with spin-orbit
interactions above, we expect the spin-smectics to emerge in a broad
range of intermediate scales from an {\em approximate}
rotationally-symmetric state, even if asymptotically crossing over to
more conventional ordered states.

\subsection{Frustrated \texorpdfstring{$J_1-J_2$}{J1-J2} spin model\label{sec:lattice model}}
Before we summarize our main findings, we discuss the frustrated
$J_1-J_2$ lattice model, studied at a microscopic level in
Refs.~\cite{bergmanOrderbydisorderSpiralSpinliquid2007,mulderSpiralOrderDisorder2010}. The
long-scale phenomenology that emerges from this model motivates our
current study.

Its Hamiltonian is given by
\begin{align}\label{eq:H_J1_J2}
    H = J_1\sum_{\langle i j\rangle}\vec{S}_{i}\cdot\vec{S}_{j} + J_2\sum_{\langle\langle i j\rangle\rangle}\vec{S}_{i}\cdot\vec{S}_{j},
\end{align}
where $\vec{S}_i$ is an $N$-component spin on site $i$ (with $N=1,2,3$
respectively corresponding to the Ising, XY, and Heisenberg models),
$\langle ij\rangle$ and $\langle\langle ij\rangle\rangle$ denote the
nearest-neighbor (NN) and next-nearest-neighbor (NNN) pairs of
sites. The frustration of the system is then induced by the
antiferromagnetic NNN interactions, $J_2 > 0$, while the sign of the
NN exchange interactions is not important (as it is nonfrustrating and can be changed by a bipartite transformation), but for concreteness and convenience we take to be $J_1 > 0$. Then,
the spin frustration is characterized by the ratio, $J_2/J_1>0$. In
the classical $S=\infty$ limit the model (\ref{eq:H_J1_J2}) exhibits
ground state degeneracy. Taking the diamond-lattice antiferromagnets
(as realized in AB$_2$X$_4$ compounds with A a magnetic ion living on
a diamond lattice and B living on a pyrochlore lattice, studied by
Bergman et al. \cite{bergmanOrderbydisorderSpiralSpinliquid2007}) for
example, for weak frustration, $0 < J_2/J_1 < 1/8$, the ground state
is the N$\acute{\text{e}}$el state. For intermediate frustration,
$1/8 < J_2/J_1 < 1/4$, the ground state is an incommensurate helical
spin-density wave that is degenerate with respect to orientation of
the ordering wavevector ${\bf q}_0$ on the so-called spiral
codimension one surface around the $\Gamma$ point. For
$1/4 < J_2/J_1$, the spiral surface exhibits open topology along the
$(111)$ axis, and in the limit $J_2/J_1\to\infty$, collapses into one
dimensional lines that correspond to the nearest-neighbor-coupled
face-centered cubic antiferromagnet.

As a consequence of the large classical ground-state degeneracy, the
ordering temperature, $T_c$, is small relative to the Curie-Weiss
exchange scale, $\Theta_{CW}$ \cite{1994AnRMS..24..453R}. For example, spinel compounds like
$\text{CoAl}_\text{2}\text{O}_\text{4}$
\cite{tristanGeometricFrustrationCubic2005,suzukiMeltingAntiferromagneticOrdering2007} and $\text{MnSc}_\text{2}\text{S}_\text{4}$
\cite{fritschSpinOrbitalFrustration2004} have
$|\Theta_{CW}| > 10-20 T_c$ and $|\Theta_{CW}|\approx 10 T_c$,
respectively. This is considered as empirical signatures of highly
frustrated magnets. As an aside, this then leads to a broad regime of
spiral classical spin liquid for temperatures $T_c<T< |\Theta_{CW}|$,
where the system thermally explores many different low-energy
configurations on the spiral surface and thereby exhibit anomalous
physical properties. This is in contrast with an even more exotic
quantum spin liquid that survives down to zero temperature
\cite{balentsSpinLiquidsFrustrated2010a}.

At low temperatures, $T<T_c$, the ordering of such magnets is
associated with the lifting of the spiral surface degeneracy that is
sensitive to the degeneracy-breaking perturbations like spin-orbit and
crystalline symmetry breaking anisotropies. In the absence of such
perturbations, the spiral surface degeneracy is lifted via quantum and
thermal fluctuations in the free energy, which select a set of
wavevectors (whose magnitude is given by the radius of spiral surface,
$q_0$) of the ordered states that one expects to be along the
crystalline symmetry axes -- the so-called
order-by-disorder \cite{villainOrderEffectDisorder1980,henleyOrderingDueDisorder1989}. The resulting
low-temperature ordered phases range from the nematic
\cite{mulderSpiralOrderDisorder2010} to a variety of spin-density-wave
states at specific wavevector
\cite{bergmanOrderbydisorderSpiralSpinliquid2007}. The ordering of the latter
self-consistently introduces a stabilizing stiffness of a conventional n$\sigma$m and thereby
determines other physical observables such as the specific
heat and structure factor \cite{bergmanOrderbydisorderSpiralSpinliquid2007,hsiehHelicalSuperfluidFrustrated2022}. As
a result, the fluctuation-generated stiffness is small, subdominant to
the higher order gradient elasticity over a large regime (set by
order-by-disorder scale), within which the system exhibits the softer
(than conventional spin-density-wave states) smectic-like elasticity
\cite{hsiehHelicalSuperfluidFrustrated2022}. Our goal here is to study
the universal low-energy description and phenomenology of such soft,
unidirectionally ordered spin-density-wave states, that we refer to as
spin-smectics. To this end, in this paper we primarily focus on an
isotropic (i.e., neglecting the order-by-disorder and lattice symmetry
breaking effects) field theory controlled by the ordering on the
spiral momentum surface.

\subsection{Results}
\label{sec:results}

\begin{figure}[t!]
\includegraphics[width=.45\textwidth]{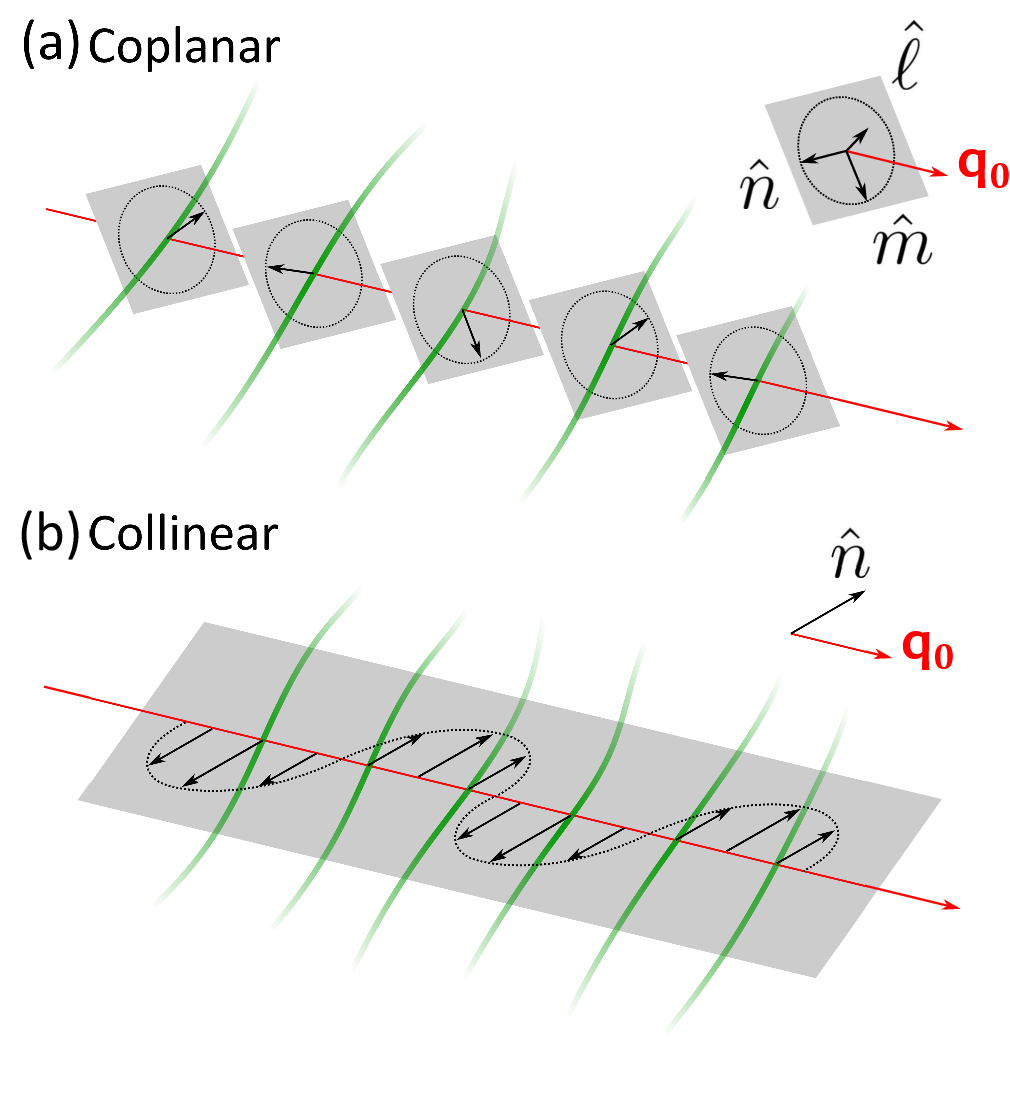}
\caption{A schematic of $N=3$ (a) coplanar and (b) collinear
  spin-smectics. The black and red arrows denote the spins $\Sv(\xv)$ and the
  spiral wavevector ${\bf q}_0$, respectively. The green (online) curves denote the spin-smectic phase fronts of constant spin magnitude and orientation, with the nearby fronts separated by an arbitrarily chosen phase of $2\pi/3$. Upper right insets: The
  coplanar state is described by an orthonormal triad $\nh(\xv)$,
  $\mh(\xv)$, $\lh(\xv)$ with $\nh$ and $\mh$ defining the spin
  $\Sv(\xv)$ plane that in the absence of spin-orbit interaction (that
  is our focus here) is decoupled from ${\bf q}_0$. The collinear
  state is characterized by a unit vector $\nh(\xv)$ that denotes the
  axis of the collinear spin-density wave.}
\label{fig:SDW}
\end{figure}

\begin{table}[t!]
{\setcellgapes{1.5ex}\makegapedcells
    \centering
    \begin{tabular}{c|c|c}
       & Coplanar & Collinear 
       \\ \hline
       $N_{GM}$ & $2N-3$ & $N$ 
       \\ \hline
       $N=1$ & N/A & $Sm$ \\ \hline
       $N=2$ & $Sm$ & $Sm$ + $XY$ \\ \hline
       $N\ge 3$ & soft n$\sigma$m, (\ref{eq:H_coplanar_O(N)_sim}) & $Sm$ + $O(N)$, (\ref{eq:H_collinear_O(N)_sim}) \\ \hline
    \end{tabular}}
  \caption{Goldstone mode number and type in the coplanar and
    collinear $O(N)$ spin-smectics for different number $N$ of spin
    components.  $N_{GM}$ is the number of the Goldstone modes. ``$Sm$'',
    ``$XY$'' and ``$O(N)$'' stand for smectic, $XY$, and $O(N)$ Goldstone
    mode models, respectively.}\label{table:GM theory}
\end{table}

\subsubsection{Coplanar smectic \texorpdfstring{$\sigma$}{sigma}-model}
Motivated by the above physical systems and the frustrated $J_1-J_2$
Heisenberg model
\cite{bergmanOrderbydisorderSpiralSpinliquid2007,mulderSpiralOrderDisorder2010},
generalized to $N$ spin components, our key result (summarized for
different cases of interest in Table~\ref{table:GM theory}) is the
derivation of $O(N)$ smectic $\sigma$-model for
$N$-component orthonormal $\nh({\bf x})-\mh({\bf x})$ diad, described
by the Hamiltonian density
\begin{align}\label{eq:H_coplanar_O(N)_sim}
    \mathcal{H}_{coplanar}[\nh,\mh] = J\left|\nabla^2\hat{\psi}+2iq_0\partial_\parallel\hat{\psi}\right|^2 + \kappa(\nabla \hat{\hat{L}})^2\ .
\end{align}
It describes the nonlinear Goldstone modes of the $N\ge 3$ {\em
  coplanar} spiral state at a wavevector ${\bf q}_0$ [see
Fig.~{\ref{fig:SDW}}(a)]. In the above, we defined components of spin
fluctuations
$\hat{\hat{L}}_{\alpha\beta} = n_\alpha m_\beta - n_\beta m_\alpha$
transverse to the spiral plane $\nh-\mh$ and the complex vector
$\hat{\psi} = (\nh + i\mh)/\sqrt{2}$. For simplicity of presentation
here, we neglected inessential (quantitative) anisotropy in $\kappa$,
discussed in the main body of the paper. Throughout this paper, we use
the subscripts $\parallel$ ($\equiv\hat{\bf q}$) and $\perp$ to denote the
axes parallel and perpendicular to ${\bf q}_0$, respectively. The $J$
modulus describes the ``soft'' Goldstone-mode elasticity for the
fluctuations in the $\nh-\mh$ plane, which for weak excitations reduces to
the standard smectic form, (\ref{eq:H_sm_O(1)}), and also introduces $\kappa_\parallel$ stiffness for $\hat{\hat{L}}$. The $\kappa$ terms
give linear-gradient elasticity for the out-of-$\nh-\mh$-plane
fluctuations. We find that $B,K,\kappa_\parallel\propto S_0^2$,
$\kappa_\perp\propto S_0^4$, predicting a divergent anisotropy near
the critical point, where $S_0\to 0$.

For the special case of $N=3$, the spin-smectic $\sigma$-model in
(\ref{eq:H_coplanar_O(N)_sim}) reduces to
\begin{align}
    \mathcal{H}_{coplanar} =&\ J\left|\nabla^2\hat{\psi}+2iq_0\partial_\parallel\hat{\psi}\right|^2 + \kappa(\nabla \lh)^2
    \nonumber\\
    =&\  \frac{J}{2}\left(\nabla^2\hat{n}-2q_0\partial_\parallel\hat{m}\right)^2+\frac{J}{2}\left(\nabla^2\hat{m}+2q_0\partial_\parallel\hat{n}\right)^2
    \nonumber\\
    &\ + \kappa(\nabla \lh)^2,
\end{align}
where the complex $3$-vector $\hat{\psi} = (\nh + i\mh)/\sqrt{2}$,
defining an orthonormal triad $\nh\times\mh = \lh$.

Easy-plane (XY) anisotropy $-(\lh\cdot\hat{c})^2$ locks $\lh$ along the anisotropy axis $\hat{c}$, reducing the model to a $N=2$ conventional smectic
  $\sigma$-model, of e.g., cholesteric and Fulde-Ferrell states for $\phi$ with
  $\nh = (\cos\phi, \sin\phi)$.

\subsubsection{Collinear smectic \texorpdfstring{$\sigma$}{sigma}-model}
For {\em collinear} spiral states [see Fig.~{\ref{fig:SDW}}(b)], the
low-energy Goldstone modes are a smectic-like pseudo-phonon
$u({\bf x})$ corresponding to the spin-density wave phase and a unit
vector $\nh({\bf x})$ that describes local spin orientation. These are
characterized by a Hamiltonian density
\begin{align}\label{eq:H_collinear_O(N)_sim}
    \mathcal{H}_{collinear}[u,\nh] =&\ Bu_{qq}^2 + K(\nabla^2 u)^2 + \kappa(\nabla\hat{n})^2,
\end{align}
where $u_{qq} = \partial_q u + (\nabla u)^2/2$ is the
rotationally-invariant strain tensor. In the presence of easy-axis
anisotropy $g(\nh\cdot\hat{c})^2$ common to magnetic crystalline materials, when $g<0$, the spin
orientation $\nh$ freezes out, leading to low-energy $N=1$ smectic
$\sigma$-model described by a single smectic phonon derived in
Sec.~\ref{N1smectic}. When $g>0$, the spin
orientation $\nh$ is locked perpendicular to $\hat{c}$, resulting in an $N=2$ collinear smectic, such as the Larkin-Ovchinnikov state.

\subsubsection{Thermal fluctuations and structure factor}

Focusing on the physical case, $N=d=3$, we show that both the coplanar
and collinear states are described by a smectic phonon and two $XY$
Goldstone modes at the harmonic level, i.e., at the Gaussian fixed point, neglecting effects of nonlinearities that may lead to a crossover to a nontrivial spin-smectic fixed point, thereby modifying these predictions at long
scales. Accordingly, they both exhibit quasi-long-range and
long-range orders in their spin-density modulation and spin
orientation orders, respectively. As illustrated in Fig.~\ref{fig:Sq_schematic}, this then leads in 3d to anisotropic
double-power-law peaks in their static spin structure factor at
$\pm n{\bf q}_0$ ($n = 1, 2, 3, ...$) [see Eq.~(\ref{eq:structure_fac}) for a more detailed
form],
\begin{align}
    \mathcal{S}({\bf q}) \approx \sum_{n}P_n({\bf q}-n{\bf q}_0) + \sum_{n}P_n({\bf q}+n{\bf q}_0),
\end{align}
where ($a=1$)
\begin{align}\label{eq:peak_q}
    P_n({\bf k}_\perp=0,k_\parallel) \sim \frac{1}{|k_\parallel|^{2-\eta_n}} + \frac{T}{\kappa}\frac{1}{|k_\parallel|^{1-\eta_n+\frac{1}{1+2\eta_n}}}
\end{align}
with the positive temperature-dependent exponent
$\eta_n = n^2 q_0^2 T/16\pi\sqrt{BK}$. In the above, the first term is
the leading (narrower) smectic Goldstone mode contribution of the
power-law peak, while the second term is the sub-leading (broader)
contribution from the $XY$ Goldstone mode fluctuations, with the
prediction valid at small $T/\kappa$,  expected to become
important as this ratio becomes of order $1$ or larger.

We discuss the effects of various symmetry-breaking perturbations that
exist in real materials, including the lattice anisotropy and
spin-orbit coupling. The former breaks the $O(d=3)$ spatial rotational
symmetry, and thereby leads to a crossover for the smectic phonon mode to XY correlation (see Fig.~\ref{fig:Csm}). This results in true Bragg
peaks (delta functions), as in conventional long-range-ordered states,
but may keep the power-law tail at intermediate scales if the
symmetry-breaking perturbation is weak. On the other hand, the
spin-orbit coupling, as e.g., Dzyaloshinskii-Moriya (DM) interaction
in a helimagnet (or a chiral twist term in a cholesteric), breaks
$O(N=3)\times O(d=3)$ symmetries down to the diagonal subgroup. For
the coplanar state, this locks the orientation of $\lh$ perpendicular
to ${\bf q}_0$, and thereby freezes out the two XY Goldstone modes,
resulting in single-power-law peak with the second term in
(\ref{eq:peak_q}) suppressed.

\subsubsection{Transition into the spin-smectic: \texorpdfstring{$O(N)$}{O(N)} de Gennes
  model}
\begin{figure}[t!]
\includegraphics[width=.45\textwidth]{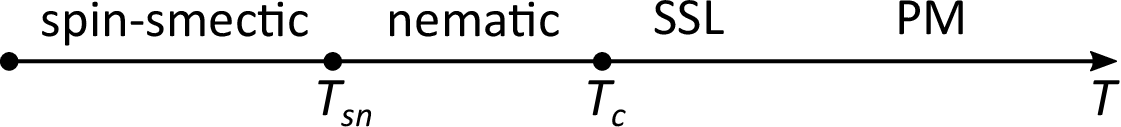}
\caption{Schematic phase diagram for the frustrated $J_1-J_2$
  Heisenberg model with low-temperature spin-density wave that also
  breaks (at intermediate scales for a lattice system with weak
  spin-orbit coupling) translational and spatial rotational
  symmetries, thereby exhibiting what we dub a ``spin-smectic''.  We
  expect it to melt at $T_{sn}$ into an orientationally-ordered
  spin-nematic (both in spin, $\langle S_\alpha S_\beta\rangle\nsim \delta_{\alpha\beta}$, and coordinate, $\langle q_i q_j\rangle\nsim \delta_{ij}$, spaces). The spin-nematic
  then disordered at $T_c$ into a spiral spin liquid (SSL)
  (characterized by a short-range smectic order) that then crosses over into
  a paramagnetic state (PM).}\label{fig:phase_diagram}
\end{figure}

To describe the criticality associated with the phase transition into
these spin-smectic states, illustrated in
Fig.~\ref{fig:phase_diagram}, we also derive a generalized $O(N)$ de
Gennes model \cite{gennesPhysicsLiquidCrystals1995}, with Hamiltonian
density given by,
\begin{align}
    \mathcal{H}_{GL} =&\ r|\vec{\psi}|^2 + v_1|\vec{\psi}|^4 + \frac{v_2}{2}|\vec{\psi}\cdot\vec{\psi}|^2 + \frac{J}{2}\left|(i\nabla-q_0\delta{\bf N})\vec{\psi}\right|^2
    \nonumber\\
    &\ + K_s(\nabla\cdot\delta{\bf N})^2 + K_{tb}(\nabla\times\delta{\bf N})^2,
\end{align}
where the complex vector $\vec{\psi} = \vec{n} + i\vec{m}$ ($\vec{n}$
and $\vec{m}$ are independent $N$-vectors, physical cases corresponding to $N=1,2,3$) describes the spin-density
wave order parameter, and $\delta{\bf N} = {\bf N} - \hat{\bf q}$ with
${\bf N}$ a {\em spatial} unit vector that describes the
translationally invariant nematic liquid that spin-smectic melts into
for $r > 0$. This Ginzburg-Landau theory, which we dub $O(N)$ spin-de
Gennes model, at low-temperatures for $r<0$ predicts the collinear
($v_2<0$) and coplanar ($v_2>0$) states discussed above.

\subsubsection{Quantum dynamics}
By including the Wess-Zumino-Witten Berry phase, that encodes the spin
precessional dynamics and corresponding spin commutation relations, we
supplement above $O(N)$ classical smectic $\sigma$-models with quantum
dynamics in the spin-coherent path integral.  For $N=3$ coplanar state
we find that Berry phase action for the smectic $\sigma$-model is
given by a
\begin{align}
    S_B =&\ \gamma\int_{\xv,t}
\left[(\partial_t\nh)^2+(\partial_t\mh)^2 
 + 2(\mh\cdot\partial_t\nh)^2\right],
\end{align}
where $\gamma$ is the uniform ferromagnetic susceptibility in the
coplanar phase.

\begin{figure}[t!]
\includegraphics[width=.4\textwidth]{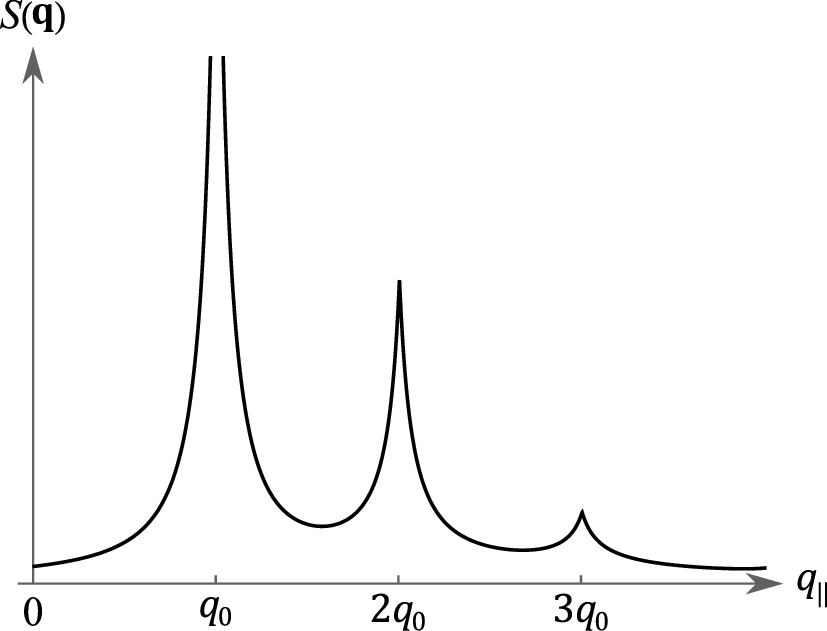}
\caption{Schematic plot of the structure factor $\mathcal{S}({\bf q})$ for the $O(3)$ collinear and coplanar spin-smectic states, that display double-power-law quasi-Bragg (as oppose to single-power-law and delta-function) peaks at ${\bf q} = \pm n{\bf q}_0$.}\label{fig:Sq_schematic}
\end{figure}

The remainder of the paper is organized as follows. In Section
\ref{sec:field theory}, we derive the continuum field theory of the
lattice model (\ref{eq:H_J1_J2}) with generalized $O(N)$ spin symmetry
and the underlying spatial $O(d)$ rotational and $T(d)$ translational
symmetries. It gives our key result, a new class of ``soft''
$O(N)\times O(d)$ $\sigma$-models that describe the universal
long-wavelength properties of spin-smectics. We show that these reduce
to the fully {\em nonlinear} Goldstone-mode field theories for the
conventional smectic liquid crystals ($N=1$), the putative FFLO paired
superfluids in imbalanced degenerate atomic gases ($N=2$), and a new
class of soft-spin-density waves ($N\ge 3$), among other physical
realizations. In Section \ref{sec:fluctuations}, we analyze the
spin-smectic in the presence of weak thermal fluctuations, assess its
stability, and compute its correlation functions, focusing on the
structure factor that exhibits crossovers across a range of length
scales. In Section \ref{sec:H_GL}, we introduce a complementary
Ginzburg-Landau model that gives spin-smectic as its ordered state and
describes the transition from the orientationally ordered to the
spin-smectic states. In Section \ref{sec:QDynamics}, starting with the
WZW term we derive the quantum dynamics for the $O(N)$ spin-smectic
$\sigma$-model.  We conclude in Sec.~\ref{sec:conclusion} with a
summary of our results 
and a discussion of open directions.

\section{Field theory of \texorpdfstring{$O(N)$}{O(N)} smectics}\label{sec:field theory}
The key feature of the frustrated lattice model, (\ref{eq:H_J1_J2}),
is its degenerate ground-state manifold (neglecting
order-from-disorder effects), defined by a spiral surface
$\varepsilon({\bf q})=\text{const.}$, where ${\bf q}$ is the wavevector
of the spin spiral state. Motivated by this class of models and corresponding
experimental realizations of frustrated helimagnets \cite{tristanGeometricFrustrationCubic2005,suzukiMeltingAntiferromagneticOrdering2007,fritschSpinOrbitalFrustration2004,gaoSpiralSpinLiquid2022,bordelonFrustratedHeisenbergEnsuremath2021}, we develop the field theory of low-energy excitations of helical
states that emerge from an {\em isotropic} spiral surface,
$|{\bf q}|=q_0$, neglecting spin-orbit coupling (SOC) and
rotational-symmetry-breaking lattice effects. To this end, we develop
an $O(N)$ Goldstone-mode $\sigma$-model by starting with an $N$-component spin $\vec{S}$ model with Hamiltonian density (throughout this paper, we employ the Einstein convention, where repeated
  indices are implicitly summed over),
\begin{align}\label{eq:H_field}
  \mathcal{H} =&\ \frac{J}{2}\left[(\nabla^2\vec{S})^2-2\bar{q}_0^2(\nabla\vec{S})^2\right] + \frac{r}{2}\vec{S}^2 + \lambda_1\vec{S}^4
                 \nonumber\\
               &+ \lambda_2(\vec{S}\times\nabla\vec{S})^2 +
                 \lambda_3(\partial_i\vec{S}\cdot\partial_j\vec{S})^2 +\ldots\;,
\end{align}
that encodes the coplanar and collinear spiral states on a $O(d)$
hyper-spherical spiral surface,
$\varepsilon(|{\bf q}|) = (q^2 - q_0^2)^2 = \text{const.}$, that can be
derived from the microscopic Hamiltonian (\ref{eq:H_J1_J2}).
We note that the Hamiltonian (\ref{eq:H_field}) respects
$O(d)\times O(N)$-rotational (and $T(d)$-translational with $d$ the
spatial dimension) symmetry, neglecting any SOC,
i.e., forbidding the inner product between the spatial ($\nabla$) and
``spin'' ($\vec{S}$) degrees of freedom.

Driven by the $J$ term, that for $r < 0$ clearly exhibits a nonzero
momentum $|{\bf q}| = \bar{q}_0$ spin condensation on an
$O(d)$-symmetric spiral surface \footnote{Generally, the wavevector
  that minimizes the energy $|{\bf q}| = q_0\neq\bar{q}_0$, as it can
  be corrected by other (than $J$) gradient terms and higher-order
  fluctuations.}, we consider (as we will show, most
strongly-fluctuating) unidirectional, single $\qv$ spin states \footnote{We do not consider mutli-$\bf q$, nor non-Landau spin liquid
  states \cite{balentsSpinLiquidsFrustrated2010a}. Determining the true ground state among the single-$\bf q$ and these states is a challenging microscopic question that falls outside of effective field theory approach taken here and is best addressed numerically.}
\begin{align}\label{eq:OP_sdw}
    \vec{S}({\bf x}) =&\ \vec{n}({\bf x})\cos({\bf q}\cdot{\bf x}) - \vec{m}({\bf x})\sin({\bf q}\cdot{\bf x}) 
    \nonumber\\
    =&\ \text{Re}\left[\vec{\psi}({\bf x})e^{i{\bf q}\cdot{\bf x}}\right],
\end{align}
where $\vec{n}$ and $\vec{m}$ are $N$-component real vector fields,
that can be compactly written as a complex $N$-vector field
\begin{align}\label{eq:S_q}
    \vec{\psi}=\vec{n}+i\vec{m},
\end{align}
with $2N$ degrees of freedom.

The remaining quadratic and quartic terms in (\ref{eq:H_field}) for
$r < 0$ encode Landau ordering and determine the amplitude of the
spin-density-wave spiral. We note that in contrast to conventional
Landau theories, here the ordering is at a nonzero wavevector, with
$\vec{S}$ condensing at a nonzero momentum $|{\bf q}|=q_0$. Therefore,
the higher derivative $\lambda_{2,3}$ terms play an important role, as
they constitute the lowest-order form needed for a generic
Goldstone-mode description of the spin-smectic $\sigma$-model
\footnote{On this account, higher order gradient quartic-in-$\vec{S}$
  terms can be included, but add nothing new beyond those already
  included in (\ref{eq:H_field}), which is the minimum form required
  for a generic description, as we verify a posteriori.}. As we will
show, the $\lambda_{2,3}$ in (\ref{eq:H_field}) are crucial for the
stabilization of the $N\ge 2$ spin-density waves.

With the ansatz (\ref{eq:OP_sdw}), the mean-field (constant part,
after dropping the fast-oscillating contributions, that on spatial
integration average out to zero) energy density is given by
\begin{align}\label{eq:H_Landau}
    \mathcal{H}_{Landau} =& \frac{\tilde{r}}{4}|\vec{\psi}|^2 +
                            \frac{v_1}{4}|\vec{\psi}|^4 +
                            \frac{v_2}{8}|\vec{\psi}\cdot\vec{\psi}|^2\;,
\end{align}
where the zeroth-order couplings (corrected by fluctuations) are given
by
\begin{align}
    \tilde{r} =&\ Jq^4 - 2Jq^2\bar{q}_0^2 + r,
    \nonumber\\
    v_1=&\ \lambda_1+\lambda_2 q^2+\lambda_3 q^4,
    \nonumber\\
    v_2=&\ \lambda_1-2\lambda_2 q^2 + \lambda_3 q^4.
\end{align}
At low temperatures, $\tilde{r}<0$ (and $v_1>0$), the spin-density wave is ordered and can be parametrized as
\begin{align}\label{eq:psi_parametrized}
    \vec{\psi} = S_0\left(\nh\cos\chi + i\mh\sin\chi\right)\;,
\end{align}
where $S_0=|\vec{\psi}|$ and $\chi$ are, respectively, the overall and
relative amplitudes of order parameters $\vec{n}$ and $\vec{m}$. The
(mean-field) magnitude of the ordering wavevector ${\bf q}$ is then
determined by minimizing the Landau free energy (\ref{eq:H_Landau}),
with direction chosen spontaneously.  For
$\lambda_2,\lambda_3\ll\lambda_1$ we find,
\begin{align}
    |{\bf q}| \equiv q_0 \approx \bar{q}_0\;.
\end{align}

It is straightforward to see that the Landau theory
(\ref{eq:H_Landau}) predicts two qualitatively distinct spin-density
waves. With the parametrization (\ref{eq:psi_parametrized}), the $v_2$
quartic interaction can be written as
\begin{align}
    |\vec{\psi}\cdot\vec{\psi}|^2 =&\ (n^2-m^2)^2 + 4(\vec{n}\cdot\vec{m})^2
    \\
    =&\ S_0^4\cos^2(2\chi) +
       S_0^4\sin^2(2\chi)(\hat{n}\cdot\hat{m})^2.
\label{eq:v2}       
\end{align}

For $v_2<0$ (or $N=1$), it is clearly minimized by
\begin{align}\label{eq:collinear_conditions}
    \hat{n}\parallel\hat{m}
\end{align}
with arbitrary phase, $\chi$. This allows us to rewrite the order parameter as
\begin{align}
    \vec{\psi}_{collinear} = S_0\nh e^{i({\bf q}_0\cdot{\bf
  x}+\chi)}\;,
  \label{psiCollinear}
\end{align}
where the minimization of (\ref{eq:H_Landau}) then gives,
\begin{align}
    S_0 = \sqrt{\frac{|\tilde{r}|}{2v_1+v_2}}\;.
\end{align}
The resulting collinear spin-density wave (\ref{eq:OP_sdw}) is then
given by oscillatory magnitude,
\begin{align}\label{eq:S^2_collinear}
    S^2_{collinear}=S_0^2\cos^2({\bf q}_0\cdot{\bf x}+\chi)\;.
\end{align}

In contrast, for $v_2>0$ (and $N>1$), minimization of (\ref{eq:v2}), then gives,
\begin{align}\label{eq:coplanar_conditions}
    \hat{n}\perp\hat{m},\quad \chi = \pi/4 + n\pi,
\end{align}
where $n$ is an arbitrary integer. This gives a coplanar, i.e.,
helical spin-density-wave state characterized by the order parameter
\begin{align}
    \vec{\psi}_{coplanar} = \frac{S_0}{\sqrt{2}}(\nh+i\mh),
\end{align}
with a constant magnitude
\begin{align}
    S^2_{coplanar} =&\ n^2\cos^2({\bf q}_0\cdot{\bf x})+m^2\sin^2({\bf q}_0\cdot{\bf x})
    \nonumber\\
    =&\ S_0^2,
\end{align}
given by
\begin{align}
    S_0 = \sqrt{\frac{|\tilde{r}|}{2v_1}}\;.
\end{align}
Importantly, because of the absence of SOC both the collinear and
coplanar states have decoupled spin orientation and wavevector
${\bf q}_0$, particularly distinguishing the coplanar state from the
helical state of a DM helimagnet and a cholesteric state of a chiral
liquid crystal.

Based on the above analysis, below we first derive $\sigma$-models for
the cases of $N=1,2$, which, as we will show correspond to the
familiar ordered states of a conventional smectic and the putative
Fulde-Ferrell-Larkin-Ovchinnikov (FFLO) superfluids, respectively.  We
then study the $N=3$ case (and its generalization to larger $N$),
corresponding to collinear and coplanar spin-density waves, described
by two distinct ``soft'' $O(3)$ nonlinear smectic
$\sigma$-models. These are relevant to the ordered states of
``codimension one'' frustrated magnetic systems reviewed in
Sec.~\ref{sec:lattice model}.

\subsection{\texorpdfstring{$N=1$}{N=1}: conventional smectic}
\label{N1smectic}
We begin with the simplest case of $N=1$, corresponding to a scalar
number density $\rho$, that according to field theory
(\ref{eq:H_field}) condenses into a {\em scalar}-density wave,
\begin{align}\label{eq:rho_q}
    \rho({\bf x}) =&\ \text{Re}\left[\psi({\bf x})e^{i{\bf q}\cdot{\bf x}}\right]
    \nonumber\\
    =&\ |\psi|\cos[ {\bf q}\cdot{\bf x} + qu({\bf x})],
\end{align}
where $\psi=|\psi|e^{iqu}$, with smectic amplitude $|\psi|$ and
$u({\bf x})$ the phonon Goldstone mode, corresponding to the
displacement along ${\bf q}$, familiar from conventional smectic
liquid crystals.  In our generalized formulation this $N=1$ state
corresponds to a necessarily collinear case, with one-component ``vectors''
$n\parallel m$, parametrized by $n=\rho\cos(qu)$ and $m=\rho\sin(qu)$.

At low temperatures in the ordered smectic state, the amplitude
$|\psi|$ is well approximated by a mean-field value
\begin{equation}
  \rho_0=\sqrt{n^2+m^2}=\sqrt{-\tilde{r}/(2v_1+v_2)}\;,
\end{equation}
with only small gapped fluctuations. Thus, with the ansatz
(\ref{eq:rho_q}) and its derivatives
\begin{align}
    \nabla\rho=&\ \rho_0\text{Re}\left[i({\bf q}+q\nabla u)e^{i{\bf q}\cdot{\bf x}+iq u}\right],
    \nonumber\\
    \nabla^2\rho=&\ \rho_0\text{Re}\left[(-({\bf q}+q\nabla u)^2+iq\nabla^2 u)e^{i{\bf q}\cdot{\bf x}+iq u}\right],
\end{align}
the model (\ref{eq:H_field}) (neglecting $\lambda_2$ and $\lambda_3$
that are unimportant for $N=1$) reduces to the familiar nonlinear
smectic Goldstone-mode elasticity,
\begin{align}
    \mathcal{H} =&\ \frac{1}{4} J\rho_0^2\left[({\bf q} + q\nabla u)^4 + q^2(\nabla^2 u)^2 - 2\bar{q}_0^2({\bf q} + q\nabla u)^2\right]
    \nonumber\\
    &\ + \frac{1}{4}r\rho_0^2 + \frac{3}{8}\lambda_1\rho_0^4 + ...
    \nonumber\\
    =&\ a(q^2-\bar{q}_0^2)u_{qq} + Bu_{qq}^2 + K(\nabla^2u)^2 +E({\bf q}),
\end{align}
where in the first equality, we dropped fast oscillating pieces, that
average away after spatial integration of the energy density. Above,
the rotationally-invariant strain tensor is
\begin{align}\label{eq:u_qq}
    u_{qq}=\hat{\bf q}\cdot\nabla u +(\nabla u)^2/2
\end{align}
and two independent elastic constants $B$ and $K$, at zeroth order
(i.e., model (\ref{eq:H_field}) dependent) are given by,
\begin{align}
    a=4K=J\rho_0^2q^2,\quad B=J\rho_0^2q^4,
\end{align}
The minimization of the constant part of $\cH$,
\begin{align}
    E({\bf q}) = \frac{1}{4}\rho_0^2(Jq^4-2Jq^2\bar{q}_0^2+r) + \frac{3}{8}\rho_0^4\lambda_1,
\end{align}
over $\rho_0$ and $q$, gives $|{\bf q}|=\bar{q}_0$, that ensures the
vanishing of the coefficient of the linear in the strain $u_{qq}$
term, and thereby guarantees the stability of the smectic state. We
note that with the inclusion of fluctuations, the optimum wavevector
gets corrected, i.e., $|{\bf q}| = q_0 \neq \bar{q}_0$, so as to
eliminate the linear in $u_{qq}$ contribution order by order (akin to
what's done with the order parameter $\rho_0$), which amounts to
ensuring that the expansion in the nonlinear strain $u_{qq}$ is done
around the correct (fluctuation-corrected) ground state. With this, by
choosing ${\bf q} = q_0\hat{\bf z}$, we then recover the familiar
nonlinear elastic, fully rotationally invariant smectic Goldstone-mode
$\sigma$-model,
\begin{align}
  \mathcal{H}_{sm} = Bu_{zz}^2 + K(\nabla^2u)^2.
  \label{eq:H_sm_O(1)}
\end{align}

We observe that the emergence of smectic elasticity
(\ref{eq:H_sm_O(1)}) is expected, based on spatial rotational symmetry
encoded in (\ref{eq:H_field}), that requires a fully rotationally
invariant nonlinear strain tensor $u_{zz}$. To see this we note that
the global rotation of the undistorted $u = 0$ smectic state is
characterized by a rotation of ${\bf q}_0$,
\begin{align}\label{eq:q0_rotation}
    q_0\hat{\bf z} \to {\bf q}'_0 = q_0\cos\theta\hat{\bf z} + q_0\sin\theta\hat{\bf x},
\end{align}
which corresponds to the phonon displacement
\begin{align}
    u_0({\bf x}) = z(\cos\theta - 1) + x \sin\theta.
\end{align}
Using $u_0({\bf x})$ inside (\ref{eq:H_sm_O(1)}), straightforward analysis then shows that
the nonlinear strain $u_{zz}$ and
thereby the energy (\ref{eq:H_sm_O(1)}) indeed vanishes under such rotation, i.e., $\cH_{sm}[u_0] = 0$.

More generally, the global rotation of the smectic state (\ref{eq:q0_rotation}) is equivalent to the following transformation of $u$:
\begin{align}
    u({\bf x}) \to u({\bf x}) + u_0({\bf x}),
\end{align}
as can be seen from its identification with the corresponding phase transformation of $\psi({\bf x})$ in Eq.~(\ref{eq:rho_q}). The latter transforms the nonlinear strain tensor
\begin{align}
    u_{zz} =&\ \partial_z u + \oh(\nabla u)^2
    \nonumber\\
    \to &\ \partial_z u + \cos\theta - 1 + \oh \left[\nabla u + (\cos\theta - 1)\hat{\bf z} + \sin\theta\hat{\bf x} \right]^2
    \nonumber\\
    =&\ (\cos\theta\hat{\bf z} + \sin\theta\hat{\bf x})\cdot\nabla u + \oh(\nabla u)^2,
\end{align}
and thereby leaves the form of Hamiltonian (\ref{eq:H_sm_O(1)}) unchanged with $\hat{\bf q} = \hat{\bf z}$ rotated to $\hat{\bf q}' = \cos\theta\hat{\bf z} + \sin\theta\hat{\bf x}$, i.e., $\cH_{sm}[u({\bf x}) + u_0({\bf x})] = \cH_{sm}[u({\bf x})]$.

\subsection{\texorpdfstring{$N=2$}{N=2}: XY smectic and FFLO/PDW superfluid}

For $N=2$, in its ordered regime the Hamiltonian (\ref{eq:H_field})
encodes a soft planar spin-density wave of XY spins. Physically, this
is relevant for frustrated magnetic systems with an easy plane
anisotropy and ``striped'' (e.g., Pair-Density Wave (PDW), FFLO, and
other) nonzero-momentum superfluids
\cite{radzihovskyFluctuationsPhaseTransitions2011a}. The latter is
characterized by a complex field, whose real and imaginary parts are
isomorphic to a two-component real vector field. The $U(1)$ symmetry
of the superfluid then maps to the $SO(2)$ spin-rotational symmetry of
the $N=2$ spin-smectic. Below, we show that the putative FF
(time-reversal-breaking, amplitude uniform) and LO
(time-reversal-preserving, amplitude modulated) PDWs states are
isomorphic to the coplanar and collinear spin-density waves, emerging
from the $N=2$ field theory, respectively.

The low-energy properties of a striped superfluid can be qualitatively
captured by the following order parameter
\begin{align}\label{eq:Delta_FFLO}
    \Delta_{FFLO}({\bf x}) = \Delta_+({\bf x}) e^{i{\bf q}\cdot{\bf x}} + \Delta_-({\bf x}) e^{-i{\bf q}\cdot{\bf x}},
\end{align}
where the two complex scalar fields,
\begin{align}
    \Delta_{\pm}({\bf x}) = \Delta_{\pm}^0({\bf x})e^{i\phi_{\pm}({\bf x})}
\end{align}
distinguishing between the coplanar and collinear (i.e., the so-called
FF and LO) states. The imaginary and real parts of the two complex
order parameters in (\ref{eq:Delta_FFLO}) can equivalently be encoded
in a spin language, via two real two-component vectors $\vec{n},\vec{m}$
via
\begin{align}\label{eq:S_FFLO}
    \vec{S}_{FFLO} = \vec{n}({\bf x})\cos({\bf q}\cdot{\bf x}) - \vec{m}({\bf x})\sin({\bf q}\cdot{\bf x}),
\end{align}
where
\begin{align}\label{eq:FFLO_mapping}
    \vec{n} =&\ (\Delta_+^R + \Delta_-^R, \Delta_+^I + \Delta_-^I),
    \nonumber\\
    \vec{m} =&\ (\Delta_+^I - \Delta_-^I, \Delta_-^R - \Delta_+^R),
\end{align}
with the superscripts $R$ and $I$ respectively denoting the real and imaginary parts of the corresponding complex fields. 

As in the analysis of the previous subsection, here too the two phases
are controlled by the sign of $v_2$. For $v_2<0$, the Landau free
energy (\ref{eq:H_Landau}) selects the collinear state that satisfies
the conditions (\ref{eq:collinear_conditions}) that together with
(\ref{eq:FFLO_mapping}) gives
$|\Delta_+| = |\Delta_-|\equiv \Delta_0$. The order parameter
(\ref{eq:Delta_FFLO}) then reduces to the familiar LO state
\begin{align}
    \Delta_{LO} = 2\Delta_0 e^{i\phi}\cos({\bf q}\cdot{\bf x}+\theta),
\end{align}
where the phases
\begin{align}
    \phi=(\phi_+ + \phi_-)/2,\quad \theta=(\phi_+ - \phi_-)/2\;,
\end{align}
are the LO superfluid phase and its smectic phonon, respectively.  In
terms of the two-component vector order parameters (\ref{eq:S_FFLO}),
this corresponds to the collinear case with
\begin{align}
    \vec{n}_{LO} =&\  2\Delta_0\cos\theta(\cos\phi,\sin\phi),
    \nonumber\\
    \vec{m}_{LO} =&\  2\Delta_0\sin\theta(\cos\phi,\sin\phi).
\end{align}
As detailed below and in
Ref.~\cite{radzihovskyFluctuationsPhaseTransitions2011a}, after
choosing the minimum $q = q_0$ and dropping constants, the
$O(N=2)_{collinear}$ $\sigma$-model (\ref{eq:H_collinear_O(N)_sim})
reduces to the Goldstone mode Hamiltonian given by the coupled smectic
and XY sectors
\cite{radzihovskyQuantumLiquidCrystals2009,radzihovskyFluctuationsPhaseTransitions2011a},
\begin{align}\label{eq:H_LO}
    \mathcal{H}_{LO} = Bu_{qq}^2 + K(\nabla^2u)^2 + \rho_s^{\parallel}(\partial_\parallel\phi)^2 + \rho_s^{\perp}(\partial_\perp\phi)^2,
\end{align}
where $u=\theta/q_0$. Notice that $\rho_s^\perp$ vanishes when the
current-current interaction, $\lambda_2\to 0$ (see below and
Ref.~\cite{radzihovskyFluctuationsPhaseTransitions2011a}). Namely, it
is required to capture the universal low-energy Goldstone-mode
energetics of the LO state.

For $v_2>0$, the Landau free energy (\ref{eq:H_Landau}) is minimized
by the coplanar state that satisfies the conditions
(\ref{eq:coplanar_conditions}), that together with
(\ref{eq:FFLO_mapping}) gives
$\Delta_+^R\Delta_-^I - \Delta_-^R\Delta_+^I= 0$ and
$\Delta_+^R\Delta_-^R + \Delta_+^I\Delta_-^I= 0$, which is equivalent
to $\Delta_+ = 0$ or $\Delta_- = 0$. Thus, the order parameter
(\ref{eq:Delta_FFLO}) reduces to the FF state
\begin{align}
    \Delta_{FF} = \Delta_0 e^{i{\bf q}\cdot{\bf x}+i\phi},
\end{align}
where $\phi=\phi_+$ and amplitude $\Delta_0 = \Delta_+^0$ is
uniform. In terms of the vector order parameter (\ref{eq:S_FFLO}),
this corresponds to the coplanar state, with orthogonal vectors
\begin{align}
  \vec{n}_{FF} =  \Delta_0(\cos\phi,\sin\phi),\
  \vec{m}_{FF} =  \Delta_0(\sin\phi,-\cos\phi)\;.
\end{align}
Thus, the Goldstone-mode $O(N=2)_{coplanar}$ $\sigma$-model is
described by a single smectic phonon (see below and
Ref.~\cite{radzihovskyFluctuationsPhaseTransitions2011a})
\begin{align}\label{eq:H_FF}
    \mathcal{H}_{FF} = Bu_{qq}^2 + K(\nabla^2u)^2,
\end{align}
where $u=\phi/q_0$.

We next turn to the discussions of the $N > 2$ coplanar $O(d)$-symmetric
density-wave, which leads to a new class of soft nonlinear $O(N > 2)$
$\sigma$-model.

\subsection{\texorpdfstring{$O(N)$}{O(N)} coplanar smectic}

For the coplanar helical state, which satisfies the condition
(\ref{eq:coplanar_conditions}) and $N\ge 2$, the order parameter can
be written as
\begin{align}\label{eq:S_coplanar}
    \vec{S} = S_0\text{Re}\left[\hat{\psi}e^{i{\bf q}\cdot{\bf x}}\right],
\end{align}
where
\begin{align}\label{eq:hat_psi_coplanar}
    \hat{\psi}({\bf x})=\frac{\hat{n}({\bf x})+i\hat{m}({\bf x})}{\sqrt{2}}
\end{align}
is a complex N-component vector field with $|\hat{\psi}|^2=1$,
described by orthonormal real vectors, $\hat{n}$ and
$\hat{m}$.

To count the number of Goldstone modes, we first note that the $O(N)$ group consists of $N(N-1)/2$ generators of rotation that correspond to independent planes in the $N$-dimensional spin space. For an ordered state that breaks the entire $O(N)$ group, there will be $N(N-1)/2$ Goldstone modes. For the coplanar state, the symmetry group of the order parameter is $O(N-2)$ (due to the subtraction of $\nh$ and $\mh$ axes). The Goldstone modes then live on $O(N)/O(N-2) = S_{N-1}\times S_{N-2}$ manifold with $2N-3$ Goldstone modes \footnote{Alternatively, we can consider $N$ orthonormal basis vectors of the spin space: $\nh$, $\mh$, and $\{\lh^\alpha\}$ ($\alpha = 1,...,N-2$). The generators of rotation that act on the orthonormal diad correspond to the $\nh-\mh$, $\nh-\lh^{\alpha}$, and $\mh-\lh^{\alpha}$ planes. This then leads to $1+(N-2)+(N-2) = 2N-3$ Goldstone modes that live on the compact
$S_{N-1}\times S_{N-2}$ manifold, with $S_{N-1}$ and $S_{N-2}$ (for a given choice) corresponding to the rotations of $\{\nh-\mh,\nh-\lh^{\alpha}\}$ and $\{\mh-\lh^{\alpha}\}$, respectively}.

To derive the Goldstone-mode classical Hamiltonian of the coplanar
state, we substitute (\ref{eq:S_coplanar}) into (\ref{eq:H_field}) and
first consider the simplest case with $\lambda_2=\lambda_3=0$, which
gives (see details in Appendix~\ref{app:gradient_terms})
\begin{align}
    \mathcal{H}_J = a(q^2-\bar{q}_0^2)\left|\nabla\hat{\psi}+i{\bf
  q}\hat{\psi}\right|^2 +
  \bar{J}\left|\nabla^2\hat{\psi}+2iq\partial_\parallel\hat{\psi}\right|^2\;,
  \label{Ha}
\end{align}
where the zeroth-order parameters above are given by
\begin{align}
    2a = 4\bar{J} = JS_0^2.
\end{align}
By selecting $|{\bf q}|\equiv q_0=\bar{q}_0$ to eliminate the $a$
(first) term in the Hamiltonian above, and expressing the result in
terms of the orthonormal triad, we obtain:
\begin{align}\label{eq:H_J_coplanar}
    \mathcal{H} = \frac{\bar{J}}{2}\left(\nabla^2\hat{n}-2q_0\partial_\parallel\hat{m}\right)^2+\frac{\bar{J}}{2}\left(\nabla^2\hat{m}+2q_0\partial_\parallel\hat{n}\right)^2.
\end{align}
We stress that this nonlinear $O(N)$ $\sigma$-model is fully
rotationally invariant for an arbitrary large spin-smectic layers
rotation $R$, corresponding to
${\bf q}_0 \to {\bf q}'_0 = R\cdot {\bf q}_0$, where in 3d
${\bf q}'_0 = q_0(\cos\theta \hat{\bf z} + \sin\theta \hat{\bf x})$
with $\hat{\bf x}$ any of the axes transverse to
$\hat{\bf q}=\hat{\bf z}$. With the definition of the order
parameter (\ref{eq:S_coplanar}), this rotation can be interpreted as
the following transformation of $\nh$ and $\mh$:
\begin{align}\label{eq:smectic_rotation}
  \nh\to &\ \nh\cos\chi_R (\xv) - \mh\sin\chi_R (\xv)\;,
           \nonumber\\
  \mh\to &\ \nh\sin\chi_R (\xv)  + \mh\cos\chi_R (\xv)\;,
\end{align}
where $\chi_R(x,z) = q_0(\cos\theta-1)z + q_0\sin\theta x$. Thus, in
the helical state the global {\em spatial} $O(d)$ rotational symmetry
of $\cH$ (\ref{eq:H_field}) maps onto an {\em inhomogeneous spin}
rotational symmetry $O_{(\nh,\mh)}$. It can be straightforwardly
verified that the transformation (\ref{eq:smectic_rotation}) leaves
the form of the Hamiltonian (\ref{eq:H_J_coplanar}) unchanged with the
$\parallel$ axis rotated to
$\hat{\bf q}' = \cos\theta \hat{\bf z} + \sin\theta \hat{\bf x}$.

\subsubsection{\texorpdfstring{$N=2$}{N=2}}

To further analyze the Hamiltonian (\ref{eq:H_J_coplanar}), we first
consider the case of $N=2$ and parametrize the orthonormal diad as
\begin{align}
    \nh_{N=2} = (\cos\chi,\sin\chi),\quad \mh_{N=2} = (-\sin\chi,\cos\chi),
\end{align}
corresponding to
\begin{align}
    \hat{\psi}_{N=2} = \frac{1}{\sqrt{2}}e^{-i\chi}(1,i),
\end{align}
where the angle $\chi$ is related to the phonon mode along ${\bf q}_0$
by $u=-\chi/q_0$. The Hamiltonian (\ref{eq:H_J_coplanar}) then
describes a smectic phonon and reduces to (\ref{eq:H_FF}) at
low-energies, where the perpendicular stiffness of
$(\nabla_\perp u)^2$ vanishes, consistent with our discussion of FF
superfluid. As we will see, for a general $N$, the soft smectic
elasticity, enforced by the underlying spatial $O(d)$ rotational
symmetry, manifests as a vanishing perpendicular stiffness,
$(\mh\cdot\nabla_\perp\nh)^2$, corresponding to {\em inhomogeneous}
spin rotations in the $\nh-\mh$ plane.

\subsubsection{\texorpdfstring{$N=3$}{N=3}}

Now we consider $N=3$ coplanar spin state. The spin space is now
spanned by the orthonormal triad $\nh$, $\mh$, $\lh$, where
\begin{align}
     \lh_\gamma = \epsilon_{\alpha\beta\gamma}\nh_\alpha\mh_\beta = \frac{1}{2}\epsilon_{\alpha\beta\gamma}\hat{\hat{L}}_{\alpha\beta}\;,
\end{align}
with
\begin{align}\label{eq:L_ab}
    \hat{\hat{L}}_{\alpha\beta} = \nh_\alpha\mh_\beta - \mh_\alpha\nh_\beta\;.
\end{align}

Notably, the $O(3)$ coplanar smectic $\sigma$-model can be expressed
in terms of the following ``spin connections'':
\begin{align}\label{eq:A_and_D_connections}
    {\bf A} = \mh\cdot\nabla\nh = i\hat{\psi}^*\cdot\nabla\hat{\psi},\quad {\bf D} = \lh\cdot\nabla\hat{\psi},
\end{align}
where ${\bf A}$ and ${\bf D}$ are real and complex spatial vector
fields, respectively. To this end, we first note that an arbitrary
vector in spin space, $\vec{v}$, can be expanded in terms of the
orthonormal triad,
\begin{align}\label{eq:nml_projection}
    \vec{v} =&\ (\nh\cdot\vec{v})\nh + (\mh\cdot\vec{v})\mh + (\lh\cdot\vec{v})\lh\;,
    \nonumber\\
    =&\ (\hat{\psi}\cdot\vec{v})\hat{\psi}^* + (\hat{\psi}^*\cdot\vec{v})\hat{\psi} + (\lh\cdot\vec{v})\lh\;.
\end{align}
This enables us to express the linear gradient term in (\ref{eq:H_J_coplanar}) as
\begin{align}\label{eq:H_J_coplanar_linear_grad}
    (\partial_\parallel\hat{m})^2 +(\partial_\parallel\hat{n})^2 =&\ 2(\hat{\bf q}\cdot {\bf A})^2 + 2|\hat{\bf q}\cdot {\bf D}|^2\;,
    \nonumber\\
    =&\ 2(\hat{m}\cdot\partial_\parallel\hat{n})^2 + (\partial_\parallel\lh)^2\;,
\end{align}
where we used
\begin{align}
  |\hat{\bf q}\cdot {\bf D}|^2 =&\ |\lh\cdot\partial_\parallel\hat{\psi}|^2 = |\hat{\psi}\cdot\partial_\parallel\lh|^2 = \oh(\partial_\parallel\lh)^2\;.
\end{align}
In the above, the ${\bf A}$ (first) and ${\bf D}$ (second) terms
correspond to the elastic moduli for the along-$\qv_0$ distortions of
in-plane (one) and out-of-plane (two) polarizations, respectively. An
important feature of the Goldstone-mode $\sigma$-model
(\ref{eq:H_J_coplanar}) is a vanishing of its transverse stiffness,
$A_\perp^2 = (\hat{m}\cdot\nabla_\perp\hat{n})^2$, guaranteed by the
underlying $O(d)$ rotational symmetry of (\ref{eq:H_field}),
corresponding to rotation of ${\bf q}$. The latter demands a vanishing
of the curvature in the transverse component to ${\bf q}_0$ in the
thermodynamic potential. The vanished stiffness then follows from the
equivalence of the {\it infinitesimal} shift
${\bf q}\to {\bf q}+\delta {\bf q}_\perp$ and the following
transformation:
\begin{align}
     {\bf A}\to{\bf A}-\delta {\bf q}_\perp + \mathcal{O}(\delta q_\perp^2).
\end{align}
In contrast, the vanishing of the transverse stiffness for the
out-of-plane polarization $\lh$, i.e., a modulus for
$|{\bf D}_{\perp}|^2 \sim (\nabla_\perp\lh)^2$ in (\ref{eq:H_J_coplanar})
is nonuniversal, unconstrained by any symmetry, and is accidental due
to a non-generic (fine-tuned) nature of (\ref{eq:H_field}) for
vanishing $\lambda_{2,3}$.

We next derive and analyze a generic coplanar $O(3)$ smectic
$\sigma$-model, by now including nonzero $\lambda_2$, $\lambda_3$. In
particular, we show that $\lambda_3$ introduces a nonzero stiffness
for $(\nabla_\perp\lh)^2$, controlling the
out-of-$\hat{n}$-$\hat{m}$-plane fluctuations, leading to our universal
Goldstone-mode $\sigma$-model of the helical state. This stiffness is
also necessary to stabilize model (\ref{eq:H_J_coplanar}), which is
otherwise unstable against thermal fluctuations in any dimension (see
Sec.~\ref{sec:fluctuations}). A complementary view on the importance
of the $\lambda_3$ coupling and the presence of
$(\partial_\perp\lh)^2$ in the Goldstone mode theory is given in
Appendix~\ref{app:H_Toner}.

To this end, we examine the contribution of nonzero $\lambda_3$ in
$\cH$ (\ref{eq:H_field}) to the $\sigma$-model in the coplanar helical
state. Relegating the details to Appendices~\ref{app:gradient_terms} and \ref{app:gauge_rep_coplanar},
using the helical order parameter (\ref{eq:S_coplanar}), dropping the
oscillatory and constant contributions, we find,
\begin{widetext}
  \begin{align}
    \label{eq:quarticStabilize}
    (\partial_i\vec{S}\cdot\partial_j\vec{S})^2 =&\ \frac{S_0^4}{4}\text{Re}\left[(\partial_i-iq_i)\hat{\psi}^*\cdot(\partial_j+iq_j)\hat{\psi}\right]^2 + \frac{S_0^4}{8}\left|\partial_i\hat{\psi}\cdot\partial_j\hat{\psi}\right|^2
    \nonumber\\
    =&\ \frac{S_0^4}{4}\text{Re}\left[D_i^*D_j + A_iA_j - q_i A_j - q_j A_i + q_i q_j\right]^2 + \frac{S_0^4}{8}\left|D_i D_j\right|^2
    \nonumber\\
    \approx &\ \frac{S_0^4}{4}\left[2|{\bf q}\cdot{\bf D}|^2 + 4({\bf q}\cdot{\bf A})^2 + 2q^2 A^2 - 4q^2({\bf q}\cdot{\bf A}) + q^4\right],
\end{align}
\end{widetext}
where in the last line, we only kept terms up to quadratic order in
${\bf A}$ and ${\bf D}$. At the minimum of the thermodynamic
potential, we remove the linear in ${\bf A}$ term with the rotationally-invariant strain tensor
\begin{align}
    \left|\nabla\hat{\psi}+i{\bf
  q}\hat{\psi}\right|^2 = \left|{\bf D}\right|^2 + ({\bf A} - {\bf q})^2,
\end{align}
which ensures the stability of the coplanar smectic state. This then leads to the correction,
\begin{align}\label{eq:lambda3_correction}
    S_0^4\left[|{\bf q}\cdot{\bf D}|^2 - q^2|{\bf D}|^2 + 4({\bf q}\cdot{\bf A})^2\right]\;,
\end{align}
with the sought-after stabilizing stiffness
\begin{align}
  |{\bf D}_\perp|^2 =&\ |{\bf D}|^2 - |\hat{\bf q}\cdot{\bf D}|^2\;, 
                       \nonumber\\
  =&\ \frac{1}{2}(\nabla_\perp\lh)^2\;.
\end{align}
Thus, by including the crucially stabilizing $\lambda_3$ contribution
to (\ref{eq:H_J_coplanar}), we now have obtained the generic form of
the $O(3)$ smectic $\sigma$-model of the helical state, as advertised in
the Results subsection of the Introduction, \ref{sec:results},
\begin{align}\label{eq:H_coplanar_O(3)}
    \mathcal{H} =&\ \bar{J}\left|\nabla^2\hat{\psi}+2iq\partial_\parallel\hat{\psi}\right|^2 + \kappa_\parallel(\partial_\parallel \lh)^2 + \kappa_\perp(\nabla_\perp \lh)^2,
\end{align}
where we included the $\kappa_\parallel$ stiffness, that, as we have
seen above, is already contained in the $\bar{J}$ contribution in
(\ref{eq:H_J_coplanar_linear_grad}) and can also be generated by other
higher-order (in $S_0$) terms.  Neglecting higher-derivative
contributions in $\bar{J}$ of (\ref{eq:H_coplanar_O(3)}) an
equivalent, smectic form of the $O(3)$ $\sigma$-model, expressed in
terms of the ``gauge'' fields is given by
\begin{align}\label{eq:H_coplanar_O(3)_lowE}
    \mathcal{H} =&\ B\left(q_0 \hat{\bf A}_\parallel-\oh\hat{\bf A}^2\right)^2 + K\left(\nabla\cdot\hat{\bf A}\right)^2
    \nonumber\\
    & + \kappa_\parallel(\partial_\parallel\lh)^2 + \kappa_\perp(\nabla_\perp\lh)^2\;,
\end{align}
where the dimensionless field (not unit vector, $\hat{\bf A}^2 \neq 1$)
\begin{align}
    \hat{\bf A} \equiv &\ {\bf A}/q_0  = \mh\cdot\nabla\nh/q_0
    \nonumber\\
    \approx &\ \nabla u,
\end{align}
leads to a conventional smectic form (\ref{eq:H_sm_O(1)}) for small
angle fluctuations. In the above, the leading contributions to the
zeroth-order parameters are
\begin{align}\label{eq:0th_parameter_coplanar}
     K,\kappa_\parallel\sim JS_0^2 q_0^2,\quad B\sim
  JS_0^2q_0^4,\quad\kappa_\perp\sim   -\lambda_3 S_0^4 q_0^2\;,
\end{align}
constrained to be $\kappa_\perp>0$ ($\lambda_3 < 0$) for stability of
the coplanar helical state.

We close this helical state derivation by noting that $\lambda_2$ term
in (\ref{eq:H_field}) gives a contribution proportional to
\begin{align}
    (\vec{S}\times\nabla\vec{S})^2 =&\ S_0^2(\nabla\vec{S})^2,
\end{align}
where we used $(\vec{S}\cdot\nabla\vec{S})^2=0$ and $S^2=S_0^2$. Since
it is proportional to an already present $(\nabla\vec{S})^2$ term in
(\ref{eq:H_field}), it simply shifts the minimum $q_0$ while
leaving the form of the resulting Goldstone mode theory
(\ref{eq:H_J_coplanar}) unchanged.

Finally, we observe that, in contrast to the stabilizing tensor
quartic operator in (\ref{eq:quarticStabilize}), a scalar quartic term,
$ (\nabla\Sv)^4$ does not introduce any new physics into the $O(3)$
helical $\sigma$-model, (\ref{eq:H_coplanar_O(3)}),
(\ref{eq:H_coplanar_O(3)_lowE}).  To see this, observe that in the
helical state,
\begin{align}
    (\nabla\Sv)^4 =&\ \frac{S_0^4}{4}\left|\nabla\hat{\psi}+i{\bf q}\hat{\psi}\right|^4 + \frac{S_0^4}{8}\left|\nabla\hat{\psi}\cdot\nabla\hat{\psi}\right|^2
    \nonumber\\
    \approx &\ \frac{S_0^4 q^2}{4}\left[2\left|\nabla\hat{\psi}+i{\bf
              q}\hat{\psi}\right|^2 +
              4|\hat{\psi}^*\cdot\partial_\parallel\hat{\psi}|^2 -
              q^2\right]\;,
\label{eq:scalarQuartic}              
\end{align}
where in the second line, we only kept the terms up to quadratic order
in $\nabla$. Thus, this contribution simply shifts the condition on
$q$ that eliminates the first term in (\ref{Ha}), only leaving the following
correction to the smectic phonon elasticity,
\begin{align}
    |\hat{\psi}^*\cdot\partial_\parallel\hat{\psi}|^2 = (\mh\cdot\partial_\parallel\nh)^2.
\end{align}
Thus, the only consequence of this scalar quartic contribution
(\ref{eq:scalarQuartic}) is to modify above zeroth-order (non-generic)
expressions for $q$, $B$ and $K$, and in particular shows that the
elastic constants $B$ and $K$ are independent.

\subsubsection{\texorpdfstring{$N>3$}{N>3}}

For $N>3$, the spin space is spanned by $N$ orthonormal vectors $\nh$,
$\mh$ and $\{\lh^\alpha\}$ ($\alpha = 1,...,N-2$). The (quadratic in
${\bf A}$ and ${\bf D}$) correction by the $\lambda_3$ term, which for
$N=3$ is given by (\ref{eq:lambda3_correction}), becomes
\begin{align}
    S_0^4\left[|{\bf q}\cdot{\bf D}^\alpha|^2 - q^2|{\bf D}^\alpha|^2 + 4({\bf q}\cdot{\bf A})^2\right]\;,
\end{align}
where ${\bf D}^\alpha = \lh^\alpha\cdot\nabla\hat{\psi}$. It now gives the following stabilizing out-of-plane contribution to
the $O(N>3)$ $\sigma$-model,
\begin{align}
    |{\bf D}_\perp^\alpha|^2 =&\ |\lh^\alpha\cdot\nabla_\perp\hat{\psi}|^2 = |\nabla_\perp\hat{\psi}|^2 - (i\hat{\psi}^*\cdot\nabla_\perp\hat{\psi})^2    
    \nonumber\\
    =&\ \frac{1}{4}(\nabla_\perp \hat{\hat{L}})^2,
\end{align}
where in the last line we expressed it in terms of $\hat{\hat{L}}$, the
components of out-of-plane fluctuations defined in (\ref{eq:L_ab}),
using the identity
\begin{align}\label{eq:dL_ab^2}
    \oh(\nabla \hat{\hat{L}}_{\alpha\beta})^2 =&\ \nabla(\nh_\alpha\mh_\beta)\nabla(\nh_\alpha\mh_\beta) - \nabla(\nh_\alpha\mh_\beta)\nabla(\nh_\beta\mh_\alpha)
    \nonumber\\
    =&\ (\nabla\nh)^2 + (\nabla\mh)^2 - 2(\mh\cdot\nabla\nh)^2
    \nonumber\\
    =&\ 2|\nabla\hat{\psi}|^2 - 2(i\hat{\psi}^*\cdot\nabla\hat{\psi})^2.
\end{align}

Thus, as advertised in the Results subsection of the Introduction,
\ref{sec:results}, the $O(N)$ smectic $\sigma$-model is given by
\begin{align}\label{eq:H_coplanar_O(N)}
    \mathcal{H} =&\ \bar{J}\left|\nabla^2\hat{\psi}+2iq\partial_\parallel\hat{\psi}\right|^2 + \kappa_\parallel(\partial_\parallel \hat{\hat{L}})^2 + \kappa_\perp(\nabla_\perp \hat{\hat{L}})^2.
\end{align}
When higher-order gradients are neglected, it reduces to a form
resembling a conventional smectic
\begin{align}\label{eq:H_coplanar_O(N)_lowE}
    \mathcal{H} =&\ B\left(q_0 \hat{\bf A}_\parallel-\oh\hat{\bf A}^2\right)^2 + K\left(\nabla\cdot\hat{{\bf A}}\right)^2
    \nonumber\\
    & + \kappa_\parallel(\partial_\parallel \hat{\hat{L}})^2 + \kappa_\perp(\nabla_\perp \hat{\hat{L}})^2.
\end{align}

\subsection{\texorpdfstring{$O(N)$}{O(N)} collinear smectic}
We next derive the Goldstone-mode smectic $\sigma$-model for the
collinear state for a general $N$. As discussed in Sec.~\ref{sec:field
  theory}, the order parameter of the collinear state can be
parametrized as
\begin{align}\label{eq:S_collinear}
    \vec{S} = S_0\text{Re}\left[\hat{\psi}({\bf x})e^{i{\bf q}\cdot{\bf x}}\right],
\end{align}
with
\begin{align}\label{eq:hat_psi_collinear}
    \hat{\psi}({\bf x}) = \hat{n}({\bf x})e^{i q u({\bf x})},
\end{align}
described by a unit ``polarization'' vector, $\hat{n}({\bf x})$, and a
phonon mode, $u({\bf x})$, with a parametrization redundancy that requires identification of $\hat{n}$ with $-\hat{n}$, which is already accounted for by $q u=\pi$. There are thus $N$ Goldstone modes
living on the $S_{N-1}\times S_1/Z_2$ compact manifold. We also note
that (in contrast to the coplanar state) the magnitude of such
linearly-polarized spin-density wave state oscillates in space, and
thus (\ref{eq:hat_psi_collinear}) corresponds to a {\em
  coarse-grained} spin-density order parameter.

We can now derive the $O(N)$ {\em collinear smectic} $\sigma$-model
using the representation (\ref{eq:S_collinear}) inside the $J$ piece
of $\cH$ in (\ref{eq:H_field}). Relegating technical details to
Appendix~\ref{app:gradient_terms}, we obtain
\begin{align}\label{eq:H_J_collinear}
  \mathcal{H}_J =&\ a(q^2-\bar{q}_0^2)\left[u_{qq}+(\nabla\hat{n})^2/2q^2\right] + Bu_{qq}^2 + K(\nabla^2 u)^2
                   \nonumber\\
                 & + \kappa_\parallel(\partial_\parallel\hat{n}+\nabla u\cdot\nabla\hat{n})^2 + \alpha(\nabla^2\hat{n})^2 + c(\nabla\hat{n})^2 u_{qq},
\end{align}
where $\partial_\parallel \equiv\partial_q = \hat{\bf q}\cdot\nabla$, the
familiar nonlinear smectic strain $u_{qq}$ is given by
(\ref{eq:u_qq}), and the zeroth-order elastic moduli are
\begin{align}
    \alpha = \frac{JS_0^2}{4},\quad a = 4K = \kappa_\parallel = c = JS_0^2q^2,\quad B = JS_0^2q^4.
\end{align}
We note that, as expected on general $O(d)$ symmetry grounds, at the
energy minimum $|{\bf q}|\equiv q_0=\bar{q}_0$, the phonon mode is
``soft'' (i.e., controlled by higher derivative elasticity), with a
vanishing transverse-to-${\bf q}$ stiffness, $(\nabla_\perp u)^2$.
This is enforced by the underlying $O(d)$ rotational invariance of
$\cH$ in (\ref{eq:H_field}) (including fluctuations), which
corresponds to an arbitrary rotation of the spontaneously chosen
${\bf q}$. This then ensures the vanishing of the
transverse-to-${\bf q}$ curvature in the thermodynamics potential at
the minimum ${\bf q}_0$, with the zero stiffness that then follows
from the equivalence of the following transformations,
\begin{align}
  {\bf q}\to {\bf q}+\delta {\bf q}_\perp \ \Leftrightarrow \  qu\to qu+\delta {\bf q}_\perp\cdot{\bf x},
\end{align}
where ${\bf q}\cdot\delta {\bf q}_\perp=0$, coming from the definition
of the order parameter in (\ref{eq:S_collinear}).

In contrast, the vanishing of the transverse stiffness of the Goldstone
mode $(\nabla_\perp\hat{n})^2$ is a purely accidental property of
the $J$ term in (\ref{eq:H_field}), and is generically expected to be
nonzero for the full $\cH$. Indeed, by including the $\lambda_2$ term
(with details given in Appendix~\ref{app:gradient_terms}), we find
\begin{align}
    (\vec{S}\times\nabla\vec{S})^2 =&\ S^2(\nabla\vec{S})^2 - (\vec{S}\cdot\nabla\vec{S})^2
    \nonumber\\
    =&\ \frac{3S_0^4}{8}(\nabla\hat{n})^2,
\end{align}
which gives a nonzero transverse $\kappa_\perp$ stiffness for $\hat{n}$
gradient deformations.  With this, and choosing
${\bf q} = q_0\hat{\bf z}$, we finally obtain the $O(N)$ {\em collinear
  smectic} $\sigma$-model for its $N$ Goldstone modes,
\begin{align}\label{eq:H_collinear}
  \mathcal{H} =&\ Bu_{zz}^2 + K(\nabla^2 u)^2
                 + c (\nabla\nh)^2 u_{zz}\nonumber\\
  & +\kappa_\parallel(\partial_z\hat{n} + \nabla
                 u\cdot\nabla\hat{n})^2
           + \kappa_\perp(\nabla_\perp\hat{n})^2,
\end{align}
where the zeroth-order stiffnesses are given by,
\begin{align}\label{eq:0th_parameter_collinear}
    K,\kappa_\parallel\sim JS_0^2q_0^2,\quad B\sim JS_0^2q_0^4,\quad \kappa_\perp\sim \lambda_2 S_0^4.
\end{align}
Neglecting symmetry-allowed nonlinearities in $u$ and $\hat{n}$,
\begin{align}\label{eq:coupling_collinear}
    (\nabla\nh)^2 u_{zz},\quad (\nabla u\cdot\nabla\nh)\partial_z\nh,\quad (\nabla u\cdot\nabla\nh)^2,
\end{align}
leads to two decoupled sectors of a conventional scalar smectic in $u$
and the standard $O(N)$ $\sigma$-model in $\nh$
(\ref{eq:H_collinear_O(N)_sim}). We leave the open question of the
effects of these coupling to a future study. Other term like
$\lambda_3$ also gives a corrections to $\kappa_\perp$ (proportional
to $JS_0^4 q_0^2$), but does not change the universal long-wavelength
form (\ref{eq:H_collinear}).  As expected, for $N=2$, the Hamiltonian
density (\ref{eq:H_collinear}) reduces to the smectic $\sigma$-model
of the LO superfluid, (\ref{eq:H_LO}), with the superfluid phase
representing the single Euler angle of $\nh$, corresponding to the
complex superfluid order parameter.

\section{Thermal fluctuations in the \texorpdfstring{$O(N=3)$}{O(3)} smectic states}\label{sec:fluctuations}

Having established the corresponding smectic coplanar and collinear
$\sigma$-models, we next analyze the thermodynamic properties of these
smectic spin-density wave states, with a focus on the physical case of
$N=3$. We limit our analysis to classical statistical mechanics at the
Gaussian fixed point, with the Goldstone-modes partition function
given by,
\begin{align}
    Z =&\ \int\mathcal{D}\hat{\psi}\mathcal{D}\hat{\psi}^* e^{-\beta \int d^dx\cH[\hat{\psi},\hat{\psi}^*]},
\end{align}
where $\beta=T^{-1}$ ($k_B=1$ throughout) and $\hat{\psi}$ given by
(\ref{eq:hat_psi_coplanar}) and (\ref{eq:hat_psi_collinear}) for the
coplanar and collinear states, respectively.

Namely, below we introduce the angular fields representation of the
Goldstone modes in the Hamiltonian density
$\cH[\hat{\psi},\hat{\psi}^*]$ for these two $O(3)$ smectic
states. Then, we examine their stability to small thermal fluctuations
within a Gaussian approximation, followed by a discussion of possible
consequences of the nonlinearities and symmetry-breaking
perturbations. We will then calculate Goldstone modes' correlation
functions that control low-energy, long-wavelength scattering and
thermodynamics.

\subsection{Angular representation of Goldstone modes\label{sec:ang_rep}}
\subsubsection{Collinear state}
As discussed in the previous sections, the $N=3$ collinear smectic
state (\ref{eq:S_collinear}) is characterized by a smectic phonon $u$
and a unit spin vector $\hat{n}$. The latter can be parameterized by
\begin{align}\label{eq:collinear_ang_rep}
    \hat{n}({\bf x}) = \cos\theta\cos\phi\eh_1 + \cos\theta\sin\phi\eh_2 + \sin\theta\eh_3,
\end{align}
where we chose an orthonormal frame $\eh_1$, $\eh_2$,
$\eh_3=\eh_1\times\eh_2$, such that for small fluctuation of these
angular Goldstone modes, $\theta({\bf x})$ and $\phi({\bf x})$,
\begin{align}
    \hat{n}({\bf x}) \approx \eh_1 + \phi\eh_2 + \theta\eh_3.
\end{align}
Neglecting higher-order gradients, the Hamiltonian density (\ref{eq:H_collinear}) is then given by
\begin{align}\label{eq:H_collinear_angles}
    \cH =&\  Bu_{qq}^2 + K(\nabla^2 u)^2 + \kappa_\parallel(\partial_\parallel\theta)^2 + \kappa_\parallel\cos^2\theta(\partial_\parallel\phi)^2
    \nonumber\\
    &\ + \kappa_\perp(\nabla_\perp\theta)^2 + \kappa_\perp\cos^2\theta(\nabla_\perp\phi)^2.
\end{align}

\subsubsection{Coplanar state}
The fluctuations of the $N=3$ coplanar state (\ref{eq:S_coplanar}) are parametrized by an
orthonormal triad, $\nh$, $\mh$, $\lh=\nh\times\mh$. In terms of the
complex vector field $\hat{\psi} = \nh+i\mh$, this can be
parameterized by
\begin{align}\label{eq:coplanar_ang_rep}
\lh=&\ \cos\theta\cos\phi\hat{e}_1+\cos\theta\sin\phi\hat{e}_2+\sin\theta\hat{e}_3,
    \nonumber\\
    \hat{\psi}=&\ \left[\left(\sin\theta\cos\phi-i\sin\phi\right)\hat{e}_1+\left(\sin\theta\sin\phi+i\cos\phi\right)\hat{e}_2\right.
    \nonumber\\
    &\quad\left.-\cos\theta\hat{e}_3\right]ie^{-i\chi}/\sqrt{2},
\end{align}
where $\theta$ and $\phi$ are Euler angles that parameterize the
orientation of $\lh$ and $\chi$ the rotation around $\lh$. In small
angles approximation, this gives,
\begin{align}
    \lh\approx &\ \hat{e}_1+\phi\hat{e}_2+\theta\hat{e}_3,
    \nonumber\\
    \nh\approx &\ \phi\hat{e}_1-\hat{e}_2-\chi\hat{e}_3,
    \nonumber\\
    \mh\approx &\ \theta\hat{e}_1+\chi\hat{e}_2-\hat{e}_3.
\end{align}
The coplanar Hamiltonian density (\ref{eq:H_coplanar_O(3)_lowE}) is then given by the same form of the collinear state Hamiltonian (\ref{eq:H_collinear_angles}), with the gradient of the smectic phonon replaced by
\begin{align}\label{eq:mh_dot_D_nh}
    \nabla u \to \hat{\bf A} = \frac{1}{q_0}\sin\theta\nabla\phi + \nabla u,
\end{align}
where the third Euler angle $\chi$ is associated
with the smectic phonon by $\chi=q_0 u$. We note that although the coplanar and collinear smectic states are quite
different, their low energy excitations only differ by the nonlinearities in $\hat{\bf A}$.

\subsubsection{Harmonic theory of \texorpdfstring{$O(3)$}{O(3)} smectic}

As discussed above, the Goldstone mode models for the $N=3$ collinear
and coplanar states arise from the spontaneous periodic ordering of
spins at wavevector ${\bf q}_0$, that leads to an order parameter
that breaks the $O(N=3)$-spin and $O(d=3)$-spatial symmetries. Because the two states only differ from each other by
the nonlinearities (\ref{eq:mh_dot_D_nh}), at the harmonic level, the
collinear and coplanar states are described by the same low-energy
Hamiltonian density,
\begin{align}\label{eq:H_gaussian}
    \cH_0 =&\ \cH_{0,sm}[u] + \cH_{0,spin}[\theta,\phi],
\end{align}
where
\begin{align}
    \cH_{0,sm}[u] =&\ B(\partial_\parallel u)^2 + K(\nabla^2 u)^2,
    \nonumber\\
    \cH_{0,spin}[\theta,\phi] =&\ \sum_{\varphi=\theta,\phi}\left[\kappa_\parallel(\partial_\parallel\varphi)^2 + \kappa_\perp(\nabla_\perp\varphi)^2\right],
\end{align}
with the zeroth-order elastic moduli given by
(\ref{eq:0th_parameter_coplanar}) and (\ref{eq:0th_parameter_collinear}) for the coplanar and collinear states, respectively. Thus, the harmonic model (\ref{eq:H_gaussian})
consists of decoupled smectic and two XY Goldstone modes.

At higher energies, however, the collinear and coplanar states acquire
distinct corrections to (\ref{eq:H_gaussian}), that can become
particularly important as near the melting transition where the ratio
of the XY moduli, $\kappa_\perp/\kappa_\parallel \sim S_0^2$, vanishes
as $S_0\to 0$.  For the collinear state, the higher-order term,
$\alpha(\nabla^2\nh)^2$, in (\ref{eq:H_J_collinear}) leads to the spin
Goldstone mode Hamiltonian
\begin{align}\label{eq:H_spin_sub_collinear}
  \cH^{collinear}_{0,spin}[\theta,\phi] =&\ \sum_{\varphi=\theta,\phi}\left[\kappa_\parallel(\partial_\parallel\phi)^2 + \kappa_\perp(\nabla_\perp\phi)^2\right] 
                                          \nonumber\\
                                        &\ + \alpha(\nabla^2 \theta)^2 + \alpha(\nabla^2 \phi)^2\;
\end{align}
with $\alpha = JS_0^2$. For  the coplanar state, the leading
correction in (\ref{eq:H_J_coplanar}) instead gives,
\begin{align}\label{eq:H_spin_sub_coplanar}
  \cH^{coplanar}_{0,spin}[\theta,\phi] =&\
                                          \bar{J}(\nabla^2\theta - 2q_0\partial_\parallel\phi)^2 + \bar{J}(\nabla^2\phi + 2q_0\partial_\parallel\theta)^2
                                          \nonumber\\
    &\ + \kappa_\perp(\nabla_\perp\theta)^2 + \kappa_\perp(\nabla_\perp\phi)^2
\end{align}
with $\bar{J} = JS_0^2/4$ and $\kappa_\parallel = 4\bar{J}q_0^2$.

\subsubsection{Symmetry-breaking
  perturbations}\label{subsection:symmetry-breaking perturbations}

Before analyzing thermal fluctuations in these $O(3)$ smectic states,
we note that in the solid state realizations there are two types of
natural symmetry-breaking perturbations on the Hamiltonian
(\ref{eq:H_field}), as we now discuss.

Firstly, in the presence of the underlying lattice that breaks
$O(d=3)$-spatial rotational symmetry but preserves $O(N=3)$-spin
symmetry, the ordering wavevector ${\bf q}_0$ will get energetically
pinned to high-symmetry crystalline axes, either in a microscopic
Hamiltonian by higher order exchange couplings or via quantum and/or
thermal order-by-disorder phenomena
\cite{bergmanOrderbydisorderSpiralSpinliquid2007,hsiehHelicalSuperfluidFrustrated2022}. This
in turn introduces a transverse (to ${\bf q}_0$) stiffness, $B_\perp$,
to the spin-density pseudo-phonon mode $u$ in the smectic sector,
leading to $\cH_{0,sm}\to\cH^{crystal}_{0,sm}$, where
\begin{align}\label{eq:H'_sm}
  \cH^{crystal}_{0,sm} &= B(\partial_\parallel u)^2 +
                        B_\perp(\partial_\perp u)^2 + K(\nabla^2
                        u)^2\;,\nonumber\\
  &\approx B(\partial_\parallel u)^2 +
                        B_\perp(\partial_\perp u)^2\;.
\end{align}

Secondly, the ever-present SOC locks the orientation
of spin to ${\bf q}$, breaking the independent $O(N=3)\times O(d=3)$
symmetry down to its diagonal subgroup, with the reduced $O(3)$ symmetry further broken by the accompanying
crystalline anisotropies. In the case of the coplanar state, this will
gap out the spin sector [$\cH_{0,spin}$ in (\ref{eq:H_gaussian})]. For ${\bf q}_0$
that is spatially incommensurate with the lattice, this will then
reduce Goldstone modes down to a single conventional XY phonon of a
standard discrete spin-density wave.

Nevertheless, in the case of weak symmetry-breaking perturbations,
based on a number of experimental realizations \cite{tristanGeometricFrustrationCubic2005,suzukiMeltingAntiferromagneticOrdering2007,fritschSpinOrbitalFrustration2004,gaoSpiralSpinLiquid2022,bordelonFrustratedHeisenbergEnsuremath2021,pfleidererPartialOrderNonFermiliquid2004,grigorievCriticalFluctuationsMnSi2005a,muhlbauerSkyrmionLatticeChiral2009,neubauerTopologicalHallEffect2009}, we expect an extended range of length scales over which our
$O(d)\times O(N)$ description will apply, but expect it to
asymptotically crossover to weakly-fluctuating behavior of
conventional spin-density waves.

\subsection{Stability}
As found in the previous subsection, at a quadratic level the
low-energy Hamiltonian densities are identical for the collinear and
coplanar spin-smectic states, given by (\ref{eq:H_gaussian}), with
decoupled Goldstone modes $u$, $\theta$, and $\phi$.  We first analyze
thermal fluctuations at this harmonic order, and then discuss the
effects of nonlinearities.

\subsubsection{Gaussian fluctuations}
The stability of the $O(3)$ smectic states is characterized by their
local Goldstone-mode thermal root-mean-squared (rms) fluctuations,
$\langle u^2 \rangle$, $\langle\theta^2 \rangle$,
$\langle\phi^2 \rangle$. The divergence of these quantities with
system size in the thermodynamic limit is a signature of the
instability of the spatial (for $\langle u^2 \rangle$) and spin (for
$\langle\theta^2 \rangle$ and $\langle\phi^2 \rangle$) orders.

We first analyze the stability of the spatially uniform component of
the magnetic order, characterized by (taking
$\kappa=\kappa_\parallel=\kappa_\perp$ for simplicity)
\begin{align}
    \langle\theta^2 \rangle = \langle\phi^2 \rangle =&\ \frac{T}{2\kappa}\int_{L^{-1}}^{a^{-1}}\frac{dq^d}{(2\pi)^d}\frac{1}{q^2}
    \nonumber\\
    \sim &\ \frac{T}{\kappa}\times\left\lbrace\begin{array}{ccc}
         & L^{2-d},\quad &d < 2,  \\
         & \ln(L/a),\quad &d = 2, \\
         & a^{2-d},\quad &d > 2,
    \end{array} \right.
\end{align}
where $L$ is the system size, $a$ is the UV cutoff length scale, $T$
is the temperature, and we neglected subordinate contributions in
$a/L \ll 1$. Thus, the spin orientational order is unstable for
$d \leq 2$. This is a manifestation in the uniform spin sector of our
system of the Hohenberg-Mermin-Wagner theorems
\cite{merminAbsenceFerromagnetismAntiferromagnetism1966,hohenbergExistenceLongRangeOrder1967},
where in a classical field theory at nonzero temperature, controlled
by a Gaussian fixed point, a continuous symmetry can only be
spontaneously-broken for $d>2$. As a side note, for
$\kappa_\perp\to 0$, the coplanar state (with Hamiltonian
(\ref{eq:H_spin_sub_coplanar}) that includes higher-order terms)
exhibits fluctuations
$\langle\theta\rangle^2=\langle\phi^2\rangle\sim\int_{\bf
  q}(q_z-q^2)^{-2}\sim L$ that diverges with system size. Thus, the
coplanar state is unstable in any dimensions without the stabilizing
modulus $\kappa_\perp$, that as we discussed above, arose from
including a nonzero $\lambda_3$ modulus.

The stability of the translational symmetry breaking is characterized
by rms fluctuations of the smectic-like phonon, given by
\begin{align}
    \langle u^2 \rangle =&\ \frac{T}{2}\int_{L_\perp^{-1}}^{a^{-1}_\perp} \frac{dq_\parallel dq_\perp^{d-1}}{(2\pi)^d}\frac{1}{Bq_\parallel^2 + Kq_\perp^4}
    \nonumber\\
    =&\ \frac{T}{4\sqrt{BK}}\int_{L_\perp^{-1}}^{a^{-1}_\perp} \frac{dq_\perp^{d-1}}{(2\pi)^{d-1}}\frac{1}{q_\perp^2}
    \nonumber\\
    \sim &\ \frac{T}{\sqrt{BK}}\times\left\lbrace\begin{array}{ccc}
         & L_\perp^{3-d},\quad &d < 3,  \\
         & \ln(L_\perp/a_\perp),\quad &d = 3, \\
         & a_\perp^{3-d},\quad &d > 3,
    \end{array} \right.
\end{align}
where $L_\perp$ and $a_\perp$ are respectively the system size and UV
lattice cutoff, transverse to ${\bf q}$, and again we only kept
leading contributions in $L_\perp \gg a_\perp$.  The lower-critical
dimension of the smectic (density-wave) order is thus given by
$d_{lc} = 3$, where the system exhibits logarithmically  diverging thermal
fluctuations \cite{gennesPhysicsLiquidCrystals1995,Caille1972}.

Thus, at the Gaussian level, we expect that the $O(3)$ smectic-like
helical states will exhibit long-range magnetic and quasi-long-range
translational orders in three dimensions. However, as emphasized in
Sec.~\ref{subsection:symmetry-breaking perturbations}, in crystalline
materials (but not in atomic gases), the presence of lattice
anisotropies and SOC introduces stabilizing elastic
moduli. These moduli give rise to conventional Goldstone modes at low
energies, characterized by $d_{lc} = 2$ and thus leading to long-range
order in three dimensions, as in conventional XY and Heisenberg
models.  However, for weak symmetry-breaking perturbations, we expect
strong smectic-like fluctuations in three dimensions extending over
long crossover length scales.

\begin{figure}[t!]
\includegraphics[width=.4\textwidth]{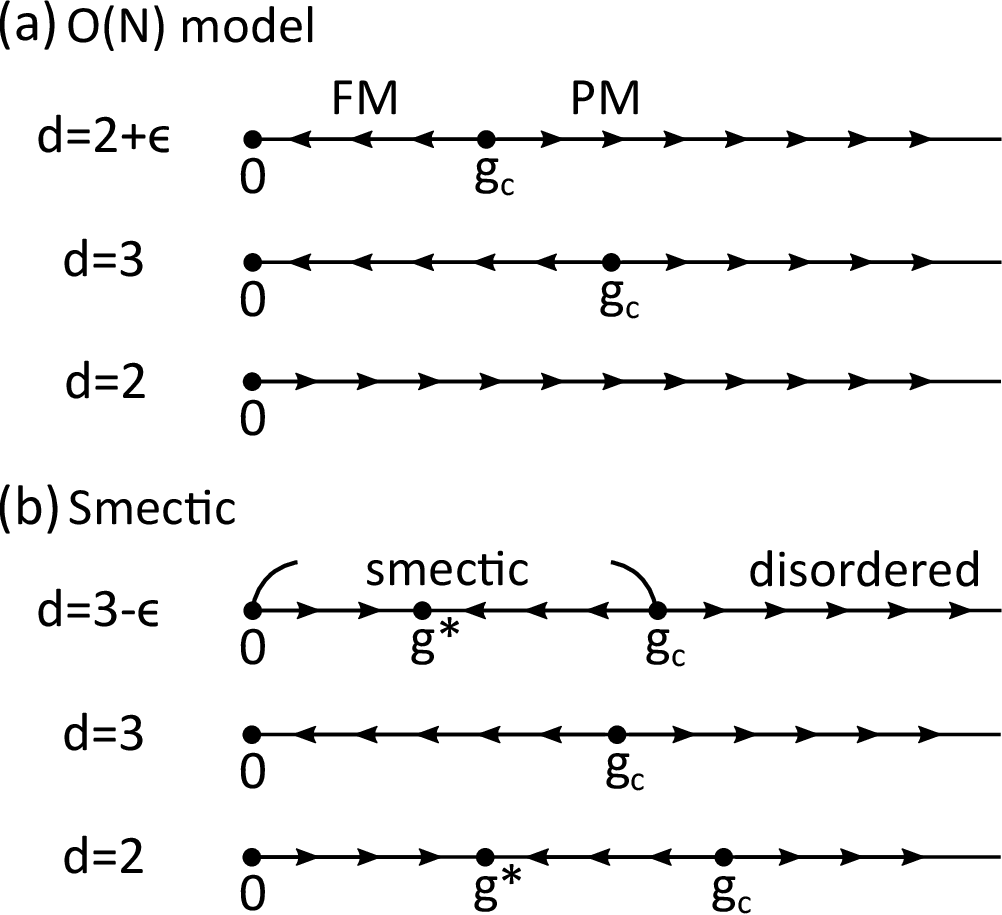}
\caption{Schematic renormalization group (RG) flows for a classical (a) ferromagnetic state in $O(N>2)$
  model and for (b) the smectic state in various dimensions. (a) The dimensionless coupling $g\sim T/\kappa$. The
  FM-PM phase transition is controlled by the {\it repulsive} critical
  point at $g=g_c\sim\epsilon/(N-2)$. For $d>2$, there is a stable
  ferromagnetic state at low temperatures, while for $d=2$, $g_c=0$,
  signifying the instability of the classical ferromagnetic state,
  destroyed by thermal fluctuations. (b) The dimensionless coupling
  $g\sim T\sqrt{B/K^3}$. At low temperatures $g<g_c$, the smectic
  state is characterized by a nontrivial infrared {\it attractive}
  fixed point at $g=g^*\sim\epsilon$, but becomes unstable below its lower-critical dimension $d_{lc} = 3$.}\label{fig:RG}
\end{figure}

\subsubsection{Nonlinearities}

As discussed above, the collinear and coplanar states are both
characterized by an $O(N=3)$ unit vector and a smectic-like phonon,
but with different symmetry-allowed nonlinear coupling terms. Below,
we will first discuss the two sectors of Goldstone modes, that have
been studied individually over the past few decades, and then will
comment on the nonlinear couplings between them.

The $O(N)$ $\sigma$-model was first studied by Polyakov
\cite{polyakovInteractionGoldstoneParticles1975}, Nelson and Pelcovits
\cite{nelsonMomentumshellRecursionRelations1977} using perturbative
renormalization group (RG) in $d = 2 + \epsilon$, which shows that for
$N>2$ the ferromagnet-paramagnet (FM-PM) transition is described by a
{\it critical} fixed point at $T_c\sim \epsilon/(N-2)$, as illustrated
in Fig.~\ref{fig:RG}(a). Consequently, the ordered state is unstable
in 2d ($T_c=0$) at any nonzero temperatures, where the correlation
length of the order parameter is finite. This is in contrast to the
$N=2$ XY model that has vanished nonlinearity, and at low temperatures
exhibits a quasi-long-range ordered Kosterlitz-Thouless phase.

The smectic Goldstone mode theory was studied by Grinstein and
Pelcovits
\cite{grinsteinAnharmonicEffectsBulk1981,grinsteinNonlinearElasticTheory1982b}
using RG for $d=3$, by Golubovi\'{c} and Wang
\cite{golubovicAnharmonicElasticitySmectics1992} in $d=2$, and by
Radzihovsky in $d=3-\epsilon$
\cite{radzihovskyFluctuationsPhaseTransitions2011a}. Remarkably, for
$d=3-\epsilon$, the smectic ordered state is described by an {\it
  attractive} fixed point at $g^*\sim\epsilon$ -- a stable critical
phase [see Fig.~\ref{fig:RG}(b)] \footnote{At the critical dimension
  $d=3$, $g^*=0$, the smectic is in principle described by a Gaussian
  fixed point, but with logarithmic corrections to the correlation
  functions and physical observables due to marginally irrelevant
  flows \cite{grinsteinAnharmonicEffectsBulk1981,grinsteinNonlinearElasticTheory1982b}
  akin to the Ising model at $d=4$.}. However, we note that these
analyses all consider pure elastic models that neglect topological
defects -- dislocations in the layered structure, which undoubtedly
unbind in 2d at any nonzero temperatures \cite{tonerSmecticCholestericRayleighBenard1981a}. Thus,
these critical phase correlations only extend out to lengths
corresponding to distance between topological defects, beyond which
the state crosses over to a translationally disordered nematic.

Now we consider the coupling between the two sectors. For the
collinear state, the leading order couplings in
(\ref{eq:coupling_collinear}) in angular representation are given by
\begin{align}
    (\nabla_\perp\nh)^2(\partial_\parallel u),\quad (\partial_\parallel\nh)\cdot(\nabla_\perp\nh\cdot\nabla_\perp u),\quad (\nabla_\perp\nh)^2(\nabla_\perp u)^2.
\end{align}
At $d=3$, the only marginal correction to the elastic moduli is
$\langle(\nabla_\perp u)^2(\nabla_\perp u)^2\rangle$ and the remaining
are irrelevant. Therefore, all coupling terms are irrelevant and at
low energies the two sectors are decoupled, described by the RG flows
in Fig.~\ref{fig:RG}, all of which are asymptotically identical to
those in
Refs.~\cite{grinsteinAnharmonicEffectsBulk1981,grinsteinNonlinearElasticTheory1982b,radzihovskyFluctuationsPhaseTransitions2011a}.
Similar argument applies to the coplanar state. However, as discussed
in Eq.~(\ref{eq:mh_dot_D_nh}), the Goldstone-mode Hamiltonian is
distinct from the collinear state by a replacement
\begin{align}
    q_0\nabla u \to\sin\theta\nabla\phi + q_0\nabla u.
\end{align}
As a result, smectic phonon $u$ and the spin field $\lh$ are
intrinsically-coupled in the low-energy Hamiltonian
(\ref{eq:H_coplanar_O(3)_lowE}). This implies that the spin field
$\lh$ can acquire universal power-law correction to its elastic
moduli, which are distinct from the $\nh$ in collinear state. We leave
the resulting RG analysis to future studies.

\subsection{Two-point correlation function of Goldstone modes}

Next we calculate the two-point correlation functions of the Goldstone
modes. This not only provides spatially resolved properties of the
system but also serves as the first step towards a calculation of the
structure factor of the following subsection. We note that as
discussed above, in principle all Goldstone modes are
coupled. However, because they are subdominant, below we neglect the
coupling between the smectic ($\chi$) and magnetic ($\theta$ and
$\phi$) sectors, a valid approximation at low energies.

\subsubsection{Smectic Goldstone mode}

Both the collinear and coplanar states are characterized by a smectic
phonon, whose correlation function is given by the following
logarithmic Caill$\acute{\text{e}}$ form \cite{Caille1972}:
\begin{align}\label{eq:C_sm}
    C_{sm}({\bf x}) \equiv &\ \langle[u({\bf x})-u(0)]^2\rangle = T\int^{\frac{1}{a}}\frac{dq_\parallel dq_\perp^2}{(2\pi)^3}\frac{1-e^{i{\bf q}\cdot{\bf x}}}{B q_\parallel^2 + K q^4}
    \nonumber\\
    \approx &\ T\int^{\frac{1}{a}}\frac{dq_\parallel dq_\perp^2}{(2\pi)^3}\frac{1-e^{i{\bf q}\cdot{\bf x}}}{B q_\parallel^2 + K q_\perp^4}
    \nonumber\\
    =&\ \frac{T}{4\pi\sqrt{BK}}\left[\ln\left(\frac{x_\perp}{a}\right) - \oh{\rm Ei}\left(\frac{-x_\perp^2}{4\lambda |x_\parallel|}\right)\right]
    \nonumber\\
    \approx &\ \left\lbrace\begin{array}{c}
         \frac{T}{4\pi\sqrt{BK}}\ln\left(\frac{x_\perp}{a}\right),\quad x_\perp\gg\sqrt{\lambda |x_\parallel|},  \\
         \frac{T}{8\pi\sqrt{BK}}\ln\left(\frac{\lambda |x_\parallel|}{a^2}\right),\quad x_\perp\ll\sqrt{\lambda |x_\parallel|},
    \end{array}\right.
\end{align}
that exhibits an anisotropic correlation at long scales, with the
coefficients in front of the $\ln$ functions differing by a factor two
in the perpendicular and parallel directions. In the above,
${\rm Ei}(x)$ is the exponential-integral function. $\lambda=\sqrt{K/B}$ is the penetration length that characterizes the
anisotropy of the smectic state.

\begin{figure}[t!]
\includegraphics[width=.4\textwidth]{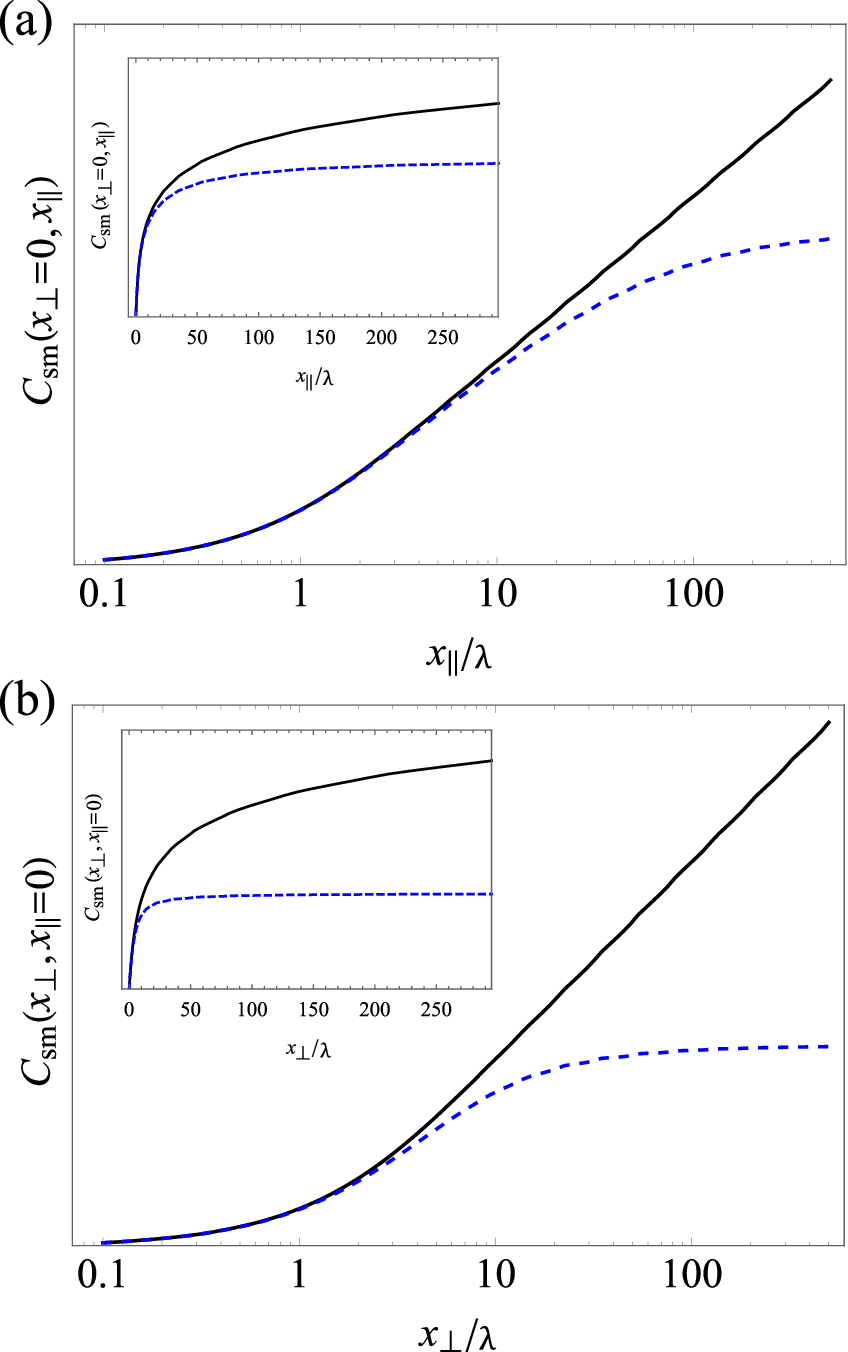}
\caption{Linear-log plot of 3d two-point correlation function of
  smectic phonon with $\lambda=\sqrt{K/B}=a$ along (a) $x_\parallel$
  and (b) $x_\perp$. The black-solid and blue-dashed curves correspond to the cases
  $\lambda_\perp=\sqrt{K/B_\perp}=\infty$ and $5\lambda$, respectively. The
  crossover scales from the smectic to XY fluctuations are given by
  $\lambda_\perp = 5\lambda$ and $\lambda_\perp^2/\lambda = 25\lambda$ in the
  perpendicular and parallel directions, respectively. Inset: same
  plot in linear scales.}\label{fig:Csm}
\end{figure}

In the presence of weak lattice anisotropy, where the spatial
rotational symmetry is explicitly broken, the smectic mode is
perturbed by a stabilizing modulus, $B_\perp\ll B$, with the
Hamiltonian given by (\ref{eq:H'_sm}). This deforms the correlation
function to be of the following form,
\begin{align}\label{eq:C_sm_crystal}
    C_{sm}^{crystal}({\bf x}) =&\ T\int_{\bf q}\frac{1 - e^{i{\bf q}\cdot{\bf x}}}{B q_\parallel^2 + B_\perp q_\perp^2 + K q^4}.
\end{align}
As illustrated in Fig.~\ref{fig:Csm}, this introduces a crossover
scale $\lambda_\perp = \sqrt{K/B_\perp}$, in 3d separating the
logarithmic ($x_\parallel < \lambda_\perp^2/\lambda$,
$x_\perp < \lambda_\perp$) and long-range-ordered
($x_\parallel > \lambda_\perp^2/\lambda$, $x_\perp > \lambda_\perp$)
regimes.

\subsubsection{Spin Goldstone modes}

For the spin sector, the Goldstone modes $\varphi = \theta,\phi$ at
low energies are described by the XY model form, with their
correlation function given by
\begin{align}\label{eq:C_xy}
    C_{xy,\varphi}({\bf x}) =&\ \langle[\varphi({\bf x})-\varphi(0)]^2\rangle
    \nonumber\\
    =&\ T\int\frac{dq_\parallel dq_\perp^2}{(2\pi)^3}\frac{1-e^{i{\bf q}\cdot{\bf x}}}{\kappa_\parallel q_\parallel^2 + \kappa_\perp q_\perp^2}
    \nonumber\\
    =&\ \frac{T}{2\pi^2\kappa_\parallel^{1/2}\kappa_\perp}\int_0^{1/\tilde{x}(a)}dp\left(1-\frac{\sin(p\tilde{x})}{p\tilde{x}}\right)
    \nonumber\\
    \approx &\ \frac{T}{2\pi^2\kappa_\parallel^{1/2}\kappa_\perp}\left(\frac{1}{\tilde{x}(a)}-\frac{1}{\tilde{x}({\bf x})}\right),
\end{align}
where
\begin{align}
    \tilde{x}({\bf x}) = \sqrt{\frac{x_\parallel^2}{\kappa_\parallel} + \frac{x_\perp^2}{\kappa_\perp}},\quad \tilde{x}(a) = a\sqrt{\kappa_\parallel^{-1}+\kappa_\perp^{-1}}.
\end{align}
The correlator consists of constant and power-law parts, that, as
discussed below gives rise to two power-law contributions to the peaks
of the static structure factor.

As discussed in Sec.~\ref{subsection:symmetry-breaking perturbations},
in the presence of SOC that locks the spins
perpendicular to ${\bf q}_0$, the magnetic Goldstone modes can be
pinned with a gap $k_p^2$, leading to
\begin{align}\label{eq:C_xy_soc}
    C^{soc}_{xy}({\bf x}) =&\ T\int\frac{dq_\parallel dq_\perp^2}{(2\pi)^3}\frac{1-e^{i{\bf q}\cdot{\bf x}}}{\kappa_\parallel q_\parallel^2 + \kappa_\perp q_\perp^2 + k_p^2}
    \nonumber\\
    =&\ \frac{T}{2\pi^2\kappa_\parallel^{1/2}\kappa_\perp}\int_0^{\frac{1}{\tilde{x}(a)}}\frac{dp}{1+k_p^2/p^2}\left(1-\frac{\sin(p\tilde{x})}{p\tilde{x}}\right)
    \nonumber\\
    \approx &\ \frac{T}{2\pi^2\kappa_\parallel^{1/2}\kappa_\perp}\left(\frac{e^{-k_p\tilde{x}(a)}}{\tilde{x}(a)}-\frac{e^{-k_p\tilde{x}({\bf x})}}{\tilde{x}({\bf x})}\right),
\end{align}
where the gap introduces crossover length scales $\xi^{soc}_{\parallel/\perp} = \sqrt{\kappa_{\parallel/\perp}}/k_p$ for the parallel/perpendicular directions, beyond which the spin fluctuations are suppressed.

\subsection{Structure factor}\label{sec:observable}

\begin{figure}[t!]
\includegraphics[width=.4\textwidth]{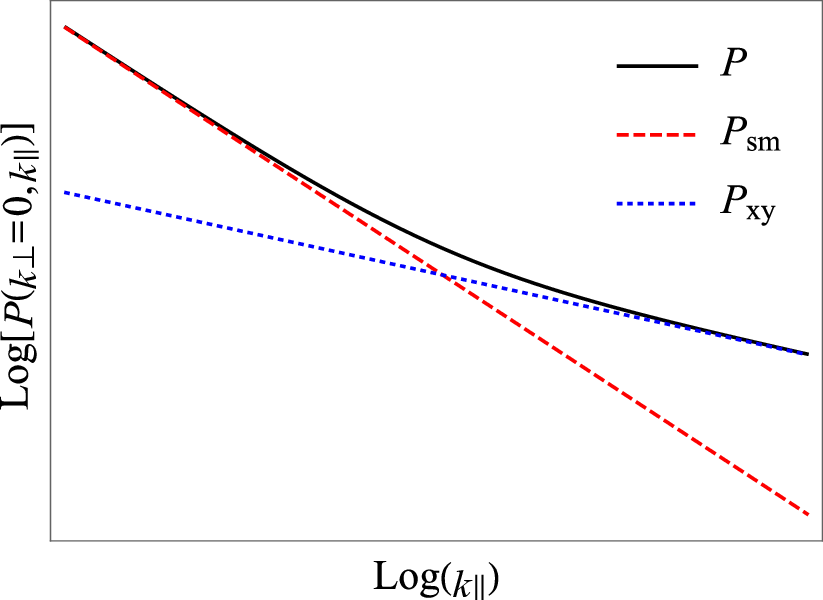}
\caption{Schematic plot of the ideal double-power-law peak $P({\bf k}) = P_{sm}({\bf k}) + P_{xy}({\bf k})$ in the structure factor along $k_\parallel$. The combined effects of smectic (red-dashed) and XY (blue-dotted) fluctuations give rise to the double-power-law peak (black-solid).}\label{fig:Sq_DP}
\end{figure}

\begin{figure}[t!]
\includegraphics[width=.45\textwidth]{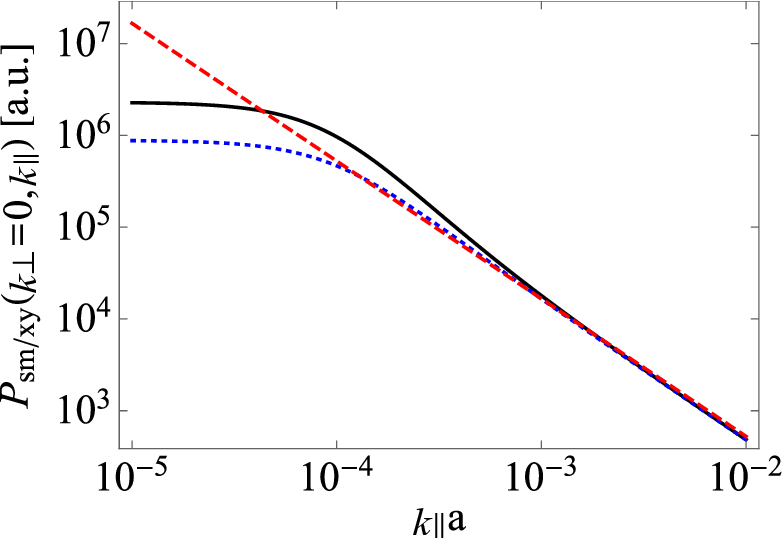}
\caption{Schematic plot of the asymptotic behavior of the power-law
  contribution in the structure factor, $P_{sm}({\bf k})$ or
  $P_{xy}({\bf k})$, along $k_\parallel$. The red-dashed line shows
  the ideal power-law behavior of $P_{sm/xy}$ with an exponent
  $1.5$. The blue-dotted curve shows the peak perturbed by an infrared
  cutoff $\xi$ (chosen to be $10000 a$, modeled by an exponential cutoff)
  due to a finite linear domain size of the system ($\xi^{cr}$, for $P_{sm}$ or
  $P_{xy}$) or SOC effects ($\xi^{soc}$, only for $P_{xy}$). The black-solid curve
  shows the power-law peak $P_{sm}$ perturbed by (in addition to the
  finite-size cutoff) the lattice anisotropy effects that lead to a
  crossover scale (chosen to be $\xi/2$) to the XY
  fluctuations.}\label{fig:Sq_perturb}
\end{figure}

The static spin structure factor is an important experimental characterization of magnetic ordering. Theoretically, this is proportional to equal-time spin-spin correlation function,
\begin{align}\label{eq:structure_fun_def}
    \mathcal{S}({\bf q}) =&\ \frac{1}{V}\int_{{\bf x},{\bf x}'}e^{i{\bf q}\cdot({\bf x}-{\bf x'})}\langle \vec{S}({\bf x})\cdot\vec{S}({\bf x'})\rangle,
\end{align}
where $V$ is the volume of the system. Below, we calculate the
structure factor for the soft spin-density waves using the Gaussian
theory (\ref{eq:H_gaussian}), focusing on its asymptotic
long-wavelength behaviors and the effects of symmetry-breaking
perturbations that are relevant in real materials. As detailed in
Appendix~\ref{app:structure_fac}, at the Gaussian level, the
asymptotic behavior of the structure factor for the collinear and
coplanar states takes the same form:
\begin{align}\label{eq:structure_fac}
    \mathcal{S}({\bf q}) \sim &\ \int_{{\bf x}}\left[e^{i({\bf q}-{\bf q}_0)\cdot{\bf x}}+e^{i({\bf q}+{\bf q}_0)\cdot{\bf x}}\right]\text{Tr}D({\bf x})e^{-\oh q_0^2 C_{sm}({\bf x})}
    \nonumber\\
    =&\ P_0({\bf q}) + P({\bf q}-{\bf q}_0)+P({\bf q}+{\bf q}_0)
\end{align}
with
\begin{align}
    P({\bf k}) = P_{sm}({\bf k})+P_{xy}({\bf k}),
\end{align}
where we retained only the fundamental $\pm {\bf q}_0$
quasi-Bragg peaks, with real spin-smectic also displaying higher harmonics $\pm n {\bf q}_0$ ($n = 2, 3, ...$). The matrix $D({\bf x})$ above is the magnetic sector correlators
defined and calculated in Appendix~\ref{app:structure_fac}, with the
trace given by
\begin{align}\label{eq:Tr_D}
    \text{Tr}D({\bf x}) = \left\lbrace\begin{array}{cc}
        e^{-\oh C_{xy,\theta}({\bf x})-\oh C_{xy,\phi}({\bf x})},\quad ({\rm collinear}). \\
        e^{-\frac{1}{4}C_{xy,\theta}({\bf x})-\frac{1}{4}C_{xy,\phi}({\bf x})},\quad ({\rm coplanar}).
    \end{array}\right.
\end{align}
The key characteristic feature of $\mathcal{S}({\bf q})$ is the 3d
quasi-Bragg peak around the ordering wavevector $\bf q_0$.  In the
above, $P_0$ is the contributions from short-range correlations that
depends smoothly on ${\bf q}$, $P_{sm}({\bf k})$ is the leading-order
singular part due to the quasi-long-range correlation of the smectic
phonon, given by ($a=1$)
\begin{align}
    P_{sm}({\bf k}) \sim  D_0\left\lbrace\begin{array}{cc}
   \frac{1}{|{\bf k}_\perp|^{4-2\eta}}, & \text{for }k_\parallel = 0,   \\
   \frac{1}{|k_\parallel|^{2-\eta}}, & \text{for }k_\perp = 0,
    \end{array}\right.
\end{align}
with $D_0 = \text{Tr}D(|{\bf x}|\to\infty)$ the Debye–Waller factor, and $P_{xy}({\bf k})$ is sub-leading singular contribution coming from the power-law dependence in $\text{Tr}D({\bf x})$, (\ref{eq:Tr_D}), which (together with the smectic correlation) is given by
\begin{align}\label{eq:P_xy}
     P_{xy}({\bf k}) \sim  \frac{T}{\kappa}\left\lbrace\begin{array}{cc}
   \frac{1}{|{\bf k}_\perp|^{2(1-\eta)+\frac{\eta}{1+\eta}}}, & \text{for }k_\parallel = 0,   \\
   \frac{1}{|k_\parallel|^{1-\eta+\frac{1}{1+2\eta}}}, & \text{for }k_\perp = 0,
    \end{array}\right.
\end{align}
where for simplicity we chose $\kappa=\kappa_\parallel=\kappa_\perp$ and the nonuniversal temperature-dependent exponent is given by,
\begin{align}\label{eq:eta}
    \eta = \frac{q_0^2 T}{16\pi\sqrt{BK}}.
\end{align}

As illustrated in Fig.~\ref{fig:Sq_DP}, the structure factor exhibits
anisotropic ``double-power-law'' peaks at ${\bf q} = \pm {\bf q}_0$
with exponents $4-2\eta$ and $2-\eta$ perpendicular and parallel to
the ordering wavevector ${\bf q}_0$, respectively.  Away from the peaks
these crossover to $2(1-\eta)+\frac{\eta}{1+\eta}$ and
$1-\eta+\frac{1}{1+2\eta}$, as dominated by the sub-leading (but
broader) contributions of $P_{xy}$. The magnitudes of the power-law
functions $P_{sm}$ and $P_{xy}$ are nonuniversal, depending on the
detailed Goldstone mode dispersions and temperature. In real systems,
it may be hard to distinguish the power-law tail of $P_{xy}$ from the
nonuniversal contribution $P_0$. However, the former may still
manifest at high temperatures or small stiffness $\kappa$ [see
Eq.~(\ref{eq:P_xy})], not only due to the enhanced magnitude of
$P_{xy}$, but also the suppression of $P_{sm}$ by the Debye-Waller
factor.

In real crystalline materials, the spin-density waves generally consist of higher harmonics that will also give rise to double-power-law peaks at ${\bf q} = \pm n{\bf q}_0$ with modified exponents $\eta \to \eta_n = n^2\eta$, see Eq.~(\ref{eq:peak_q}) and Fig.~\ref{fig:Sq_schematic}. Furthermore, the spin-density waves tend to form
domains with their wavevectors pinned by the underlying lattice. In
this case, the structure factor is given by the average of those
domain contributions, leading to peaks located on the high symmetry
axes of the Brillouin zone. As illustrated in
Fig.~\ref{fig:Sq_perturb}, this also leads to finite-size effects that
broaden the power-law contributions $P_{sm}$ and $P_{xy}$, with the
leading singular part $P_{sm}$ by widths $1/\xi^{cr}$ and
$(a_0 \xi^{cr})^{-1/2}$ ($\xi^{cr}$ the averaged domain size, $a_0$ the
longest UV scale, within which the smectic dispersion starts to
deviate) in the parallel and perpendicular directions, respectively.

As shown in Eq.~(\ref{eq:C_xy_soc}), the SOC also gives rise to
similar effects on $P_{xy}$ by introducing a gap in the magnetic
Goldstone-mode correlators, that terminates the power-law dependence
around
$\xi^{soc}_{\parallel/\perp} \approx \sqrt{\kappa_{\parallel/\perp}}/k_p$. For
the cases that spins are locked perpendicular to the ${\bf q}_0$, the
collinear and coplanar states have one and two magnetic Goldstone
modes gapped, respectively. Accordingly, for the collinear state the
asymptotic behaviors of $P_{xy}$ remains the same, while for the
coplanar state $P_{xy}$ is broadened by the infrared cutoff introduced
by the gap.

In the presence of lattice anisotropy that pins the direction of
${\bf q}_0$, the smectic correlator becomes long-range-ordered, see
Fig.~\ref{fig:Csm}. This sharpens the power-law quasi-Bragg peak of
$P_{sm}$ to a delta function Bragg peak, while modifies the exponents
of $P_{xy}$ to be $2$. As illustrated in Fig.~\ref{fig:Sq_perturb},
together with the finite-size effects, this leads to an enhanced peak
strength, within the crossover momentum scale (see the black-solid and
blue-dotted curves).

Finally, we expect that near a thermal phase transition, the ratio of
the spin stiffness $\kappa_\perp/\kappa_\parallel$ to be small, which
can lead to a sizable non-singular contribution $P_0({\bf q})$ due to
strong spin-sector fluctuations, that depend on the ``sub-leading''
moduli in (\ref{eq:H_spin_sub_collinear}) and
(\ref{eq:H_spin_sub_coplanar}) for the collinear and coplanar states,
respectively. For the coplanar states, there are strong fluctuations
on the spiral surface, which may lead to anisotropic arc-shaped peaks, observed in a classical $J_1-J_2-J_3$ Heisenberg model in Ref.~\cite{glittumArcshapedStructureFactor2021}.

\section{Ginzburg-Landau model}\label{sec:H_GL}
In Sec.~\ref{sec:field theory} we constructed the spin-density
functional designed to give condensation into a spin-smectic
state. However, such functional is unable to capture the nature of a
continuous phase transition as its disordered state is an isotropic
structureless fluid. Here, we instead construct a generalized
Ginzburg-Landau ($O(N)$ generalization of de Gennes' scalar $N=1$
model for conventional smectic liquid crystal
\cite{gennesPhysicsLiquidCrystals1995}) model that gives the $O(N)$
smectic as its ordered state and has an additional virtue of
describing the continuous transition from the orientationally ordered
to the $O(N)$ smectic ordered states.

To this end, we propose the following free energy density that
incorporates the orientation order that is characterized by the
wavevector $\hat{\bf q}$
\begin{align}\label{eq:H_GL_complete}
    \mathcal{H}_{GL} =&\ r|\vec{\psi}|^2 + v_1|\vec{\psi}|^4 + \frac{v_2}{2}|\vec{\psi}\cdot\vec{\psi}|^2 + \frac{J_\parallel}{2}\big|(i\partial_\parallel-q_0\delta N_\parallel)\vec{\psi}\big|^2
    \nonumber\\
    & +\frac{J_\perp}{2}\big|(i\nabla_\perp-q_0\delta{\bf N}_\perp)\vec{\psi}\big|^2 + K_s(\nabla\cdot\delta{\bf N})^2 
    \nonumber\\
    & + K_{t}(\hat{\bf q}\cdot\nabla\times\delta{\bf N})^2 + K_{b}(\hat{\bf q}\times\nabla\times\delta{\bf N})^2.
\end{align}
In the above, 
\begin{align}
    \delta {\bf N}({\bf x}) = {\bf N}({\bf x}) - \hat{\bf q},\quad {\bf N}\cdot {\bf N} = 1,
\end{align}
are the orientational Goldstone mode fluctuations characterizing the
``nematic'' phase from which the spin-smectic emerges. As for
conventional nematics, they are characterized by the standard
Frank-Oseen free energy with the splay ($K_s$), twist ($K_t$) and bend
($K_b$) elastic moduli. The complex vector field,
$\vec{\psi}({\bf x}) = \vec{n}({\bf x})+i\vec{m}({\bf x})$, is the
slowly-varying order parameter of spin-smectic in (\ref{eq:OP_sdw})
that characterizes the strength of spin order (note that
$\vec{n}, \vec{m}$ are not unit vectors, with their amplitude growing
in the usual Landau way below the transition). As spin-smectic emerges
from an anisotropic, orientationally-ordered nematic state. the
stiffness of the order parameter are anisotropic,
$J_\parallel\neq J_\perp$.

For clarity of presentation, in the following discussion we set $K_t=K_b=K_{tb}$, with the model then
simplifying to
\begin{align}\label{eq:H_GL}
    \mathcal{H}_{GL} =&\ r|\vec{\psi}|^2 + v_1|\vec{\psi}|^4 + \frac{v_2}{2}|\vec{\psi}\cdot\vec{\psi}|^2 
    \nonumber\\
    & + \frac{J_\parallel}{2}\big|(i\partial_\parallel-q_0\delta N_\parallel)\vec{\psi}\big|^2 +\frac{J_\perp}{2}\big|(i\nabla_\perp-q_0\delta{\bf N}_\perp)\vec{\psi}\big|^2
    \nonumber\\
    &  + K_s(\nabla\cdot\delta{\bf N})^2 + K_{tb}(\nabla\times\delta{\bf N})^2,
\end{align}
which resembles the Ginzburg-Landau model of a normal-superconductor
transition with a nonzero spin-angular momentum pairing (e.g., He3),
but with an additional gauge-invariance breaking splay stiffness
replacing the Maxwell term for the vector potential.

At high temperatures, $r>0$, the complex vector field $\vec{\psi}$ is
translationally disordered, leaving the Frank-Oseen free energy that
describes the nematic-like unidirectional order of the parent liquid
state.

At low temperatures, $r<0$ (and $v_1>0$), the spin-smectic order
emerges, characterized by a nonzero order parameter,
\begin{align}
    \vec{\psi}({\bf x})=S_0\hat{\psi},
\end{align}
where for $N=1$ or $v_2<0$, the collinear state that satisfies
(\ref{eq:collinear_conditions}) is energetically preferred with
\begin{align}\label{eq:psi_collinear}
    \hat{\psi} = \hat{n}e^{i\chi},\quad S_0^2 = -r/(2v_1+v_2).
\end{align}
Instead, for $N>1$ and $v_2>0$, the coplanar state that satisfies
(\ref{eq:coplanar_conditions}) is more stable with
\begin{align}\label{eq:psi_coplanar}
    \hat{\psi} = \frac{\hat{n}+i\hat{m}}{\sqrt{2}},\quad S_0^2 = -r/2v_1.
\end{align}
The state is characterized by the coherence length,
\begin{align}
   \xi = \sqrt{\frac{J}{2|r|}},
\end{align}
which governs the spatial distortions of of the amplitude
$|\vec{\psi}|=S_0$ and thereby the size of the so-called cybotactic
clusters near the critical point, and by the orientational ``penetration'' length
\begin{align}
    \lambda_{tb} = \sqrt{\frac{2K_{tb}}{Jq_0^2S_0^2}},
\end{align}
which is the scale that twist and bend deformation can penetrate
through the soft spin-density wave.

Below, we show that at low temperatures this generalized spin-de Gennes
model reproduces all the properties of the collinear and coplanar
spin-smectic states, but in addition captures the critical properties
of the phase transition, whose beyond-mean-field treatment of critical
behavior is a challenging problem that we leave to future studies.

\subsection{Collinear state}

For the collinear state (\ref{eq:psi_collinear}), the coupling terms in (\ref{eq:H_GL}) become ($\chi=q_0u$)
\begin{align}
    |(i\partial_i-q_0\delta N_i)\hat{\psi}|^2 =&\ (\partial_i\hat{n})^2 + q_0^2(\partial_i u + \delta N_i)^2,
\end{align}
where we suspended Einstein's summation convention (no summation over $i$) and at low temperatures, deep in the collinear spin-smectic state,
the minimization of the second term above gives \footnote{This can be
  done exactly using a Lagrangian multiplier to impose the constraint
  ${\bf N}\cdot{\bf N}=1$. See, e.g.,
  Ref.~\cite{radzihovskyNonlinearSmecticElasticity2011}.},
$\delta N\ll 1$):
\begin{align}
    \delta{\bf N}_\perp = -\nabla_\perp u.
\end{align}
This leads to the following low-energy Goldstone theory that describes
the collinear state
\begin{align}
    \mathcal{H}_{GL}^{collinear} =&\ \frac{J_\parallel S_0^2q_0^2}{2}\left(\partial_\parallel u+\frac{(\nabla_\perp u)^2}{2}\right)^2 + K_s(\nabla_\perp^2 u)^2
    \nonumber\\
    & + \frac{J_\parallel S_0^2}{2}(\partial_\parallel\hat{n})^2 + \frac{J_\perp S_0^2}{2}(\nabla_\perp\hat{n})^2,
\end{align}
in agreement with our earlier analysis in (\ref{eq:H_collinear}) with the identification of the coefficients
\begin{align}
   \frac{J_{\parallel} S_0^2 q_0^2}{2} = B,\quad K_s = K,\quad  \frac{J_{\parallel/\perp} S_0^2}{2} = \kappa_{\parallel/\perp},
\end{align}
which suggests a divergent anisotropy $J_{\perp}/J_{\parallel} = \kappa_\perp/\kappa_\parallel \sim S_0^2$ near the critical point, where $S_0\to 0$. As before, deep in the phase, here we also neglected the dislocation
defects, i.e.,  $\nabla\times\nabla u=0$.

\subsection{Coplanar state}

For the coplanar spin-smectic state (\ref{eq:psi_coplanar}), the
coupling terms in (\ref{eq:H_GL}) reduce to
\begin{align}
    |(i\partial_i-q_0\delta N_i)\hat{\psi}|^2 =&\ |\partial_i\hat{\psi}|^2 - (i\hat{\psi}^*\cdot\partial_i\hat{\psi})^2
    \nonumber\\
    &\ + (i\hat{\psi}^*\cdot\partial_i\hat{\psi}-q_0\delta N_i)^2
    \nonumber\\
    =&\ \frac{1}{4}(\partial_i\hat{\hat{L}})^2 + (\hat{m}\cdot\partial_i\hat{n}-q_0\delta N_i)^2,
\end{align}
where we suspended Einstein’s summation convention and in the last line we expressed the out-of-plane fluctuations in
terms of
$\hat{\hat{L}}_{\alpha\beta} = \nh_\alpha\mh_\beta -
\mh_\alpha\nh_\beta$ using (\ref{eq:dL_ab^2}).

In the ordered state, the minimization of the second term above gives
an emergent Higgs-like mechanism locking orientational and smectic
orders according to
\begin{align}
    \delta{\bf N_\perp}= \hat{\bf A}_\perp,
\end{align}
where the dimensionless field $\hat{\bf A} = (\mh\cdot\nabla\nh)/q_0$. This then gives the Goldstone
mode Hamiltonian for the coplanar state
\begin{align}
     \mathcal{H}_{GL}^{coplanar} =&\ \frac{J_\parallel S_0^2 q_0^2}{2}\left(|\hat{\bf A}_\parallel|-\frac{1}{2}\hat{\bf A}_\perp^2\right)^2 
     \nonumber\\
     &+ K_s(\nabla\cdot\hat{\bf A}_\perp)^2 + K_{tb}(\nabla\times\hat{\bf A}_\perp)^2
     \nonumber\\
     &+ \frac{J_\parallel S_0^2}{8}(\partial_\parallel \hat{\hat{L}})^2 + \frac{J_\perp S_0^2}{8}(\nabla_\perp\hat{\hat{L}})^2,
\end{align}
which reduces to (\ref{eq:H_coplanar_O(N)_lowE}) in the
absence of dislocations,
$\nabla\times\hat{\bf A}_\perp = 0$, with
\begin{align}
    \frac{J_{\parallel} S_0^2 q_0^2}{2} = B,\quad K_s = K,\quad \frac{J_{\parallel/\perp} S_0^2}{8} = \kappa_{\parallel/\perp}.
\end{align}
This again leads to a divergent anisotropy $J_{\perp}/J_{\parallel} \sim S_0^2$ near the critical point, where $S_0\to 0$.

As noted above, in addition to reproducing the corresponding
spin-smectic states, the spin-de Gennes model faithfully
captures the spin-Nematic-Smectic phase transition that we expect to
display rich universal critical phenomenology \cite{halperinFirstOrderPhaseTransitions1974}, whose study we leave
for the future.

\section{Quantum dynamics\label{sec:QDynamics}}

So far, all of our analysis has been confined to a classical treatment
of the spin-smectic states, valid at high temperatures. However, at low
temperatures quantum dynamics becomes important, and we need to
generalize our model by extending it to include dynamics and quantize
it via, e.g., a coherent spin path-integral formulation, following the
standard derivation of the conventional antiferromagnetic
$\sigma$-model with linear
dispersion \cite{haldaneNonlinearEnsuremathSigma1988,dombreNonlinearModelsTriangular1989}.

To this end, to capture the zero-temperature dynamics and the
associated quantum fluctuations, we now introduce the spin Berry phase
that is the Wess-Zumino-Witten action $S_{WZW}$ that encodes the
$SU(2)$ ($N=3$) spin commutator algebra into the action, corresponding
to spin precession generic to all underdamped spin systems.

To extend this to an array of spins, we sum over lattice sites with
the action then given by ($\hbar=1$):
\begin{align}\label{eq:S_B}
  S_B =&\int dt\dot\phi({\bf x},t)\left[1-\cos\theta({\bf x},t)\right]\nonumber\\
  =&\int_{\bf x} S_{WZW}[\Sh({\bf x}, t, u)]
    \nonumber\\
    =&\ - s\int_{{\bf x},t}\int_0^1 du
\Sh\cdot\partial_t\Sh\times\partial_u\Sh,
\end{align}
where $s$ is the spin magnitude quantum number, $\Sh({\bf x},t,u)$ is the
coherent spin label corresponding to orientation of a spins at $\xv$,
and auxiliary time-like variable $u$ was introduced to be able to
express $S_{WZW}$ covariently in terms of $\Sh({\bf x},t,u)$, rather
than in terms of its polar ($\theta$) and azimuthal ($\phi$) angles. It is
easy to see that $S_{WZW}$ is a boundary term that after $u$ integral
gives the solid angle in (\ref{eq:S_B}), swept out by $\Sh({\bf x},t,u)$,
which quantizes spin $s$ in integer multiples of $1/2$. This also
gives the identification of $\Sh({\bf x},t,u)$ with the physical spin
according to $\Sh({\bf x},t,0)=\Sh({\bf x},t)$ and
$\Sh({\bf x},t,1)=\hat{e}$, with the latter an arbitrary reference
spin orientation.

Below, we focus on the coplanar state, expressing $\Sh$ in terms of
the zero-wavevector $\Sv_0$ (uniform, ferromagnetic part, not to be
confused with the magnitude of spin $S_0$ in previous sections) and
the nonzero wavevector (spiral part) $\Sh_q$ contributions,
\begin{align}
    \Sh =&\ (\Sv_0 + \Sh_q)/(1 + 2\Sv_0\cdot\Sh_q + \Sv_0^2)^{1/2},
\end{align}
where
\begin{align}
    \Sh_q =&\ \frac{1}{\sqrt{2}}(\psih e^{i \qv_0\cdot\xv} + \psih^* e^{-i
    \qv_0\cdot\xv})
    \nonumber\\
    =&\ \nh\cos\qv_0\cdot\xv + \mh\sin\qv_0\cdot\xv
\end{align}
and the denominator is a normalization factor that ensures
$\Sh^2=1$. The uniform component $\Sv_0$ must be included despite
being gapped in the spin-spiral state, as it encodes the conserved
magnetization and has a nontrivial commutation relation with $\Sh_q$.

We consider small ferromagnetic fluctuations $|\Sv_0| \ll 1$ such that
\begin{align}\label{eq:S0q}
    \Sh \approx &\ \Sv_{\ell} + \Sh_q,
\end{align}
where $\Sv_{\ell} = \Sv_0-(\Sv_0\cdot\Sh_q)\Sh_q$ is the components of
$\Sv_0$ perpendicular to $\Sh_q$, i.e.,
$\Sv_{\ell}\parallel\lh$. Furthermore, because the ground state is
{\em not} a ferromagnet, uniform magnetization fluctuations $\Sv_\ell$
are gapped, i.e., characterized by a Hamiltonian
\begin{align}
    H_{uniform} =
\oh\gamma^{-1}\int_\xv\Sv_\ell^2,
\end{align}
where $\gamma$ is the uniform ferromagnetic susceptibility in the
coplanar state.

With this the evolution operator is given by ($\hbar=1$)
\begin{align}
    U_t =&\ \int [d\Sh(\xv,t,u)] 
e^{i S_B - i\int dt\left[ H_{coplanar}+H_{uniform}\right]},
\end{align}
where $H_{coplanar}$ is given by (\ref{eq:H_coplanar_O(3)}). Inserting
the form (\ref{eq:S0q}) inside $S_B$ and keeping only linear terms in
the gapped uniform magnetization $\Sv_\ell$, we find
\begin{align}
    S_B =&\ -s\int_{\xv,t}\int_0^1 du
\left[\Sv_\ell\cdot\partial_t\Sh_q\times\partial_u\Sh_q\right.
 \nonumber\\
 &\ \left. +
 \Sh_q\cdot\partial_t\Sv_\ell\times\partial_u\Sh_q+\Sh_q\cdot\partial_t\Sh_q\times\partial_u\Sv_\ell\right],
\end{align}
where we dropped the term proportional to
$ \Sh_q\cdot\partial_t\Sh_q\times\partial_u\Sh_q$, that vanishes (for
smooth configurations of $\psih$) because it oscillates strongly at
wavevector ${\bf q}_0$ \footnote{This can however lead to nontrivial
  topological contributions that, e.g., for $d=1$ leads to the
  distinction between integer and half-integer spins in Heisenberg
  antiferromagnets \cite{haldaneNonlinearFieldTheory1983}}. The first
term above vanishes because it involves three vectors lying in a plane
normal to $\Sh_q$. With integration by parts, $S_B$ reduces to a
total derivative,
\begin{align}
    S_{B} =&\ -s\int_{\xv,t}\int_0^1 du\partial_u
\left(\Sv_{\ell}\cdot\Sh_q\times\partial_t\Sh_q\right)
    \nonumber\\
    =&\ -s\int_{\xv,t}\Sv_{\ell}\cdot\Sh_q\times\partial_t\Sh_q
    \nonumber\\
    =&\ -\oh s\int_{\xv,t} \Sv_{\ell}\cdot\left(\psih\times\partial_t\psih^*
+\psih^*\times\partial_t\psih\right),
\end{align}
where in the last line we dropped the oscillating terms. Integrating
over the gapped magnetization field $\Sv_{\ell}$ and substituting
$\psih = (\nh+i\mh)/\sqrt{2}$, we find
\begin{align}
    S_B =&\ \bar{\gamma}\int dt
\left(\nh\times\partial_t\nh + \mh\times\partial_t\mh\right)^2
    \nonumber\\
    =&\ \bar{\gamma}\int dt
\left[(\partial_t\nh)^2+(\partial_t\mh)^2 
 + 2(\mh\cdot\partial_t\nh)^2\right]
 \nonumber\\
    =&\ \bar{\gamma}\int dt
\left[(\partial_t\lh)^2 
 + 4(\mh\cdot\partial_t\nh)^2\right],
\end{align}
where $\bar{\gamma}=s\gamma/2$. This form is straightforwardly
generalized to $N$ spin components, and using the identity
(\ref{eq:dL_ab^2}), gives
\begin{align}
    S_B =&\ \bar{\gamma}\int dt
\left[\oh (\partial_t\hat{\hat{L}})^2 
 + 4(\mh\cdot\partial_t\nh)^2\right],
 \end{align}
 where $\hat{\hat{L}}$ is defined in (\ref{eq:L_ab}).

\section{Summary and Conclusion}\label{sec:conclusion}

Motivated by a large number of physical realizations of unidirectional
orders in liquid crystals, degenerate atomic gases, electronic and in
particular frustrated magnetic systems, here we developed a low-energy
theory of Goldstone modes, i.e., $O(N)$ spin-smectic $\sigma$-model
that describes unidirectional ``density'' waves of generalized
$N$-component smectic. We predicted two phases -- coplanar and
collinear spin-smectics that spontaneously break $O(N)$-spin and
$O(d)$-spatial rotational symmetries in addition to the translational
symmetry along ${\bf q}_0$ \footnote{Strictly speaking, the coplanar
  smectic state does {\em not} break translational symmetry. Instead
  it breaks tensor product of translations along and rotations around the
  ordering wavevector $\qv_0$ down to their diagonal subgroup, namely,
  $U_{\text rot}(1)\times U_{\text trans}(1) \rightarrow U_{\text
    diagonal}(1)$.}.

Having established two corresponding $O(N)$ smectic $\sigma$-models,
we focused on the new physical interesting case of $N=3$ and examined
spin-smectics' stability to thermal fluctuations within a harmonic
approximation. We showed that these states are characterized by a
critical dimensions $d_{c} = 3$ below which the corresponding
mean-field order is unstable, to a state with strongly fluctuating
Goldstone modes in $d\leq3$.  We briefly discussed the nonlinearities
that couple the smectic and magnetic sectors, reserving their detailed
analysis to a future study.

We then used the developed $O(N)$ $\sigma$-models to characterize
spin-smectic phases by the correlation functions of their Goldstone
modes.  In addition to asymptotic behavior the idealized system, we
also discussed the effects of weak symmetry-breaking perturbations
that exist in real materials: (i) Lattice anisotropy breaking
spatial-$O(d=3)$ rotational symmetry by pinning ${\bf q}_0$ along high
symmetry axes of the underlying lattice, that leads to a smectic to XY
model crossover for the smectic phonon mode. (ii) Spin-orbit
interaction locking the spin orientation to ${\bf q}_0$ and thereby
breaking the $O(N=3)\times O(d=3)$ rotational symmetries down to their
diagonal subgroup. In particular, for the coplanar state, the
spin-plane normal vector $\lh$ is frozen along ${\bf q}_0$, which gaps
out the two magnetic Goldstone modes, akin to cholesterics and
DM-interacting helical magnets. In materials with weak DM spin-orbit
interactions and weak lattice pinning anisotropy compared to the
Heisenberg exchange interactions, we predict that spin-smectic
$\sigma$-models will control the low-temperature Goldstone modes and
therefore will exhibit spin-smectic structure function over a large
intermediate range of length scales before asymptotically crossing
over to a conventional $\sigma$-model behavior.

Utilizing these Goldstone modes correlation functions, we computed the
static spin structure factors, focusing on their asymptotic
long-wavelength behaviors. We showed that at the harmonic level the
$N=3$ collinear and coplanar spin-smectics are characterized by the
same asymptotic form. In 3d they both exhibit double-power-law
quasi-Bragg peaks (in contrast to the usual single-power-law for
scalar smectics and delta-function Bragg peaks in conventional
magnets) at $\pm {\bf q}_0$ due to the combined effects of the
smectic- and XY- spin Goldstone mode fluctuations. In addition, we
discussed specific applications to magnetic systems, including the
effects of various symmetry-breaking perturbations and powder
averaging over different spin-smectic domains, that change the
asymptotic behaviors of the peaks. However, we expect that even in the
ideal case without such perturbations, the double-power-law
characteristic feature may be weak, and difficult to distinguish from
the non-singular short-range correlations. Yet, we found that these
novel features are enhanced near a phase transition into the
spin-nematic state, where the effective transverse to ${\bf q}_0$ spin
stiffness vanish parametrically faster than its longitudinal
counterpart. We leave the required detailed analysis near the critical
point to future studies.

We also complemented our $O(N)$ spin-smectic $\sigma$-model
development with a $O(N)$ generalization of a de Gennes-like model
that captures the spin-nematic to spin-smectic (NA) phase transition
in terms of a complex $N$-vector order parameter characterizing the
spin-smectic state. At the mean-field level it describes the phase
transition and reproduces precisely the $O(N)$ $\sigma$-models of the
planar and collinear smectic states.  However, it raises a challenging
question of true criticality of this spin-NA transition, that we leave
for future investigations.

An extension of our study to a quantum spin-smectics is another
interesting and open direction to explore. Here, by appending our
classical theory with the WZW action--spin precessional dynamics, we
derived the quantum dynamic for the coplanar spin state. At Gaussian
level it leads to one smectic-like mode (with linear and quadratic
dispersion in the parallel and perpendicular directions, respectively)
and $2N-4$ conventional spin-density wave-like Goldstone modes with
linear dispersion. This allows for the study of the dynamic structure function obtained in neutron scattering, with the analysis left for a future study. Our analysis assumes smooth
configurations of spins, and thus neglects possible topological terms
that could play a nontrivial role in the properties of
spin-smectics. We also leave more detailed studies of this to future
research.

Our study is based on a $O(d)\times O(N)$ symmetric field theory, with
the discussion of symmetry-breaking effects incorporated
phenomenologically.  A more microscopic analysis, that allows a
quantitative assessment of such symmetry breaking terms is a necessary
next step to assess smectic $\sigma$-model applicability and range of
validity in real materials.

Other future directions include but not limited to large-$N$ and RG
analyses of the $O(N)$ smectic $\sigma$-model, the effects
of topological defects, and the generalization of spin-smectic states
to different representations of the $O(N)\times O(d)$ group and to
other symmetry groups. We hope our study can stimulate future
theoretical and experimental studies in such soft spin-density waves
and their generalizations.

\section*{Acknowledgement}
LR thanks John Toner and Arun Paramekanti for enlightening discussions. This research was supported by the Simons Investigator Award to LR
from the Simons Foundation. LR thanks The Kavli Institute for
Theoretical Physics for hospitality while this manuscript was in
preparation, during Quantum Crystals and Quantum Magnetism workshops,
supported by the National Science Foundation under Grant No. NSF
PHY-1748958 and PHY-2309135.

\appendix
\widetext

\section{Analysis of the gradient terms in the \texorpdfstring{$O(N)$}{O(N)} smectic model}\label{app:gradient_terms}
Here, we evaluate all the symmetry-allowed gradient terms in the field
theory (\ref{eq:H_field}) up to quartic order in $\vec{S}$. As
discussed in Sec.~\ref{sec:field theory}, this then leads to
universal, low-energy Goldstone mode models for both the collinear
and coplanar spin-density wave states. Specifically, we will consider
the following quadratic 
\begin{align}
    (\nabla\vec{S})^2,\quad (\nabla^2\vec{S})^2
\end{align}
and quartic terms
\begin{align}
    (\nabla\vec{S})^4,\quad (\vec{S\cdot}\nabla\vec{S})^2,\quad \sum_{ij}(\partial_i\vec{S}\cdot\partial_j\vec{S})^2.
\end{align}
We note that the important $\lambda_2$ term in (\ref{eq:H_field}) is
the linear combinations of the terms above, given by
\begin{align}
    (\vec{S}\times\nabla\vec{S})^2 = S^2(\nabla\vec{S})^2 - (\vec{S}\cdot\nabla\vec{S})^2.
\end{align}

\subsection{collinear state}
For the collinear state (\ref{eq:S_collinear}) we find (${\chi = qu}$),
\begin{align}
    \partial_i\vec{S} =&\ S_0\text{Re}\left[\left(\partial_i\hat{n}+i(\partial_i\chi+q_i)\hat{n}\right)e^{i{\bf q}\cdot{\bf x}+i\chi}\right],
    \nonumber\\
    \partial_i\partial_j\vec{S} =&\ S_0\text{Re}\left[\left(\partial_i\partial_j\hat{n}-(\partial_i\chi+q_i)(\partial_j\chi+q_j)\hat{n}+i(\partial_i\chi+q_i)\partial_j\hat{n}+i(\partial_j\chi+q_j)\partial_i\hat{n}+i(\partial_i\partial_j\chi)\hat{n}\right)e^{i{\bf q}\cdot{\bf x}+i\chi}\right].
\end{align}
After dropping the oscillating (in space) contributions (that vanish
upon spatial integration)  the quadratic terms are given by
\begin{align}
    (\nabla\vec{S})^2 =&\ \frac{S_0^2}{2}\left[(\nabla\hat{n})^2 + (\nabla\chi+{\bf q})^2\right],
    \nonumber\\
    (\nabla^2\vec{S})^2 =&\ \frac{S_0^2}{2}\left\lbrace(\nabla^2\hat{n})^2 + 2(\nabla\chi+{\bf q})^2(\nabla\hat{n})^2 + (\nabla\chi+{\bf q})^4 + (\nabla^2\chi)^2 + 4[(\nabla\chi+{\bf q})\cdot\nabla\hat{n}]^2\right\rbrace,
\end{align}
where we used $\hat{n}\cdot\nabla^2\hat{n}+(\nabla\hat{n})^2=0$. The quartic terms are given by
\begin{align}
    S^2(\nabla\vec{S})^2 =&\ \frac{S_0^4}{8}\left[3(\nabla\hat{n})^2 + (\nabla\chi+{\bf q})^2\right],
    \nonumber\\
    (\vec{S}\cdot\nabla\vec{S})^2 =&\ \frac{S_0^4}{8}(\nabla\chi+{\bf q})^2,
    \nonumber\\
    (\nabla\vec{S})^4 =&\ \frac{S_0^4}{8}\left[3(\nabla\hat{n})^4 + 3(\nabla\chi+{\bf q})^4 + 2(\nabla\chi+{\bf q})^2(\nabla\hat{n})^2\right],
    \nonumber\\
    (\partial_i\vec{S}\cdot\partial_j\vec{S})^2 =&\ \frac{S_0^4}{8}\left\lbrace 3(\nabla\hat{n})^4 + 3(\nabla\chi+{\bf q})^4 + 2[(\nabla\chi+{\bf q})\cdot\nabla\hat{n}]^2 \right\rbrace.
\end{align}
We used above expression in the main text to derive the collinear
smectic $\sigma$-model.

\subsection{coplanar state}
In the coplanar spin-wave state, characterized by the order parameter
(\ref{eq:S_coplanar}) we instead find
\begin{align}
    \partial_i\vec{S} =&\ S_0\text{Re}\left[(\partial_i\hat{\psi}+iq_i\hat{\psi})e^{i{\bf q}\cdot{\bf x}}\right]\;,
    \nonumber\\
    \partial_i\partial_j\vec{S} =&\ S_0\text{Re}\left[(\partial_i\partial_j\hat{\psi}+iq_i\partial_j\hat{\psi}+iq_j\partial_i\hat{\psi}-q_iq_j\hat{\psi})e^{i{\bf q}\cdot{\bf x}}\right],
\end{align}
where $\hat{\psi}=(\hat{n}+i\hat{m})/\sqrt{2}$ with $\hat{n}\cdot\hat{m}=0$. After dropping the oscillating (in space) terms, the quadratic terms are given by
\begin{align}
    (\nabla\vec{S})^2 =&\ \frac{S_0^2}{2}\left|\nabla\hat{\psi}+i{\bf q}\hat{\psi}\right|^2 \;,
    \nonumber\\
    (\nabla^2\vec{S})^2 =&\ \frac{S_0^2}{2}\left[\left|\nabla^2\hat{\psi}+2i{\bf q}\cdot\nabla\hat{\psi}\right|^2 + 2q^2\left|\nabla\hat{\psi}+i{\bf q}\hat{\psi}\right|^2 - q^4 \right],
\end{align}
where we used the identity $\text{Re}[\hat{\psi}^*\cdot\nabla^2\hat{\psi}] = -|\nabla\hat{\psi}|^2$. The quartic terms are given by
\begin{align}
    (\nabla\vec{S})^4 =&\ \frac{S_0^4}{4}\left|\nabla\hat{\psi}+i{\bf q}\hat{\psi}\right|^4 + \frac{S_0^4}{8}\left|\nabla\hat{\psi}\cdot\nabla\hat{\psi}\right|^2 \;,
    \nonumber\\
    (\partial_i\vec{S}\cdot\partial_j\vec{S})^2 =&\ \frac{S_0^4}{4}\text{Re}\left[(\partial_i-iq_i)\hat{\psi}^*\cdot(\partial_j+iq_j)\hat{\psi}\right]^2 + \frac{S_0^4}{8}\left|\partial_i\hat{\psi}\cdot\partial_j\hat{\psi}\right|^2 \;,
\end{align}
and $(\vec{S\cdot}\nabla\vec{S})^2 = 0$ because $S^2=\text{const.}$ for the coplanar state.

In terms of $\hat{n}$ and $\hat{m}$ the expressions reduce to
\begin{align}
    \left|\nabla\hat{\psi}+i{\bf q}\hat{\psi}\right|^2 =&\ \oh \left(\nabla\hat{n}-{\bf q}\hat{m}\right)^2 + \oh \left(\nabla\hat{m}+{\bf q}\hat{n}\right)^2,
    \nonumber\\
    \left|\nabla^2\hat{\psi}+2i{\bf q}\cdot\nabla\hat{\psi}\right|^2 =&\ \oh \left(\nabla^2\hat{n}-2{\bf q}\cdot\nabla\hat{m}\right)^2 + \oh \left(\nabla^2\hat{m}+2{\bf q}\cdot\nabla\hat{n}\right)^2,
    \nonumber\\
    \left|\partial_i\hat{\psi}\cdot\partial_j\hat{\psi}\right|^2 =&\ \frac{1}{4}\left(\partial_i\hat{n}\cdot\partial_j\hat{n}-\partial_i\hat{m}\cdot\partial_j\hat{m}\right)^2 + \frac{1}{4}\left(\partial_i\hat{n}\cdot\partial_j\hat{m}+\partial_i\hat{m}\cdot\partial_j\hat{n}\right)^2,
    \nonumber\\
    \text{Re}\left[(\partial_i-iq_i)\hat{\psi}^*\cdot(\partial_j+iq_j)\hat{\psi}\right]^2 =&\ \left(\oh\partial_i\hat{n}\cdot\partial_j\hat{n} + \oh\partial_i\hat{m}\cdot\partial_j\hat{m} + q_i\mh\cdot\partial_j\nh + q_j\mh\cdot\partial_i\nh + q_i q_j\right)^2.
\end{align}

\section{Gauge field representation of \texorpdfstring{$O(3)$}{O(3)} coplanar
  \texorpdfstring{$\sigma$}{sigma}-model}\label{app:gauge_rep_coplanar}
In this appendix we reformulate the coplanar smectic $\sigma$-model
in terms of gauge (spin-connection) fields constructed from
$\nh$, $\mh$.  For $N=3$, $\hat{\psi}$, $\hat{\psi}^*$ (or
equivalently $\nh$, $\mh$) and $\lh=\nh\times\mh$ span the spin space:
\begin{align}
\vec{a}\cdot\vec{b} =&\ (\hat{\psi}^*\cdot\vec{a})(\hat{\psi}\cdot\vec{b}) + (\hat{\psi}\cdot\vec{a})(\hat{\psi}^*\cdot\vec{b}) + (\lh\cdot\vec{a})(\lh\cdot\vec{b})
\nonumber\\
=&\ (\hat{n}\cdot\vec{a})(\hat{n}\cdot\vec{b}) + (\hat{m}\cdot\vec{a})(\hat{m}\cdot\vec{b}) + (\lh\cdot\vec{a})(\lh\cdot\vec{b}).
\end{align}
With this identity, all the terms above can be written in terms of a
real vector (in space) field ${\bf A}$ and a complex vector field
${\bf D}$, defined by
\begin{align}
    {\bf A} = i\hat{\psi}^*\cdot\nabla\hat{\psi},\quad {\bf D} = \lh\cdot\nabla\hat{\psi},
\end{align}
giving
\begin{align}
    \left|\nabla\hat{\psi}+i{\bf q}\hat{\psi}\right|^2 =&\ \left|{\bf D}\right|^2 + ({\bf A} - {\bf q})^2,
    \nonumber\\
    \left|\nabla^2\hat{\psi}+2i{\bf q}\cdot\nabla\hat{\psi}\right|^2 =&\ 
    \left|{\bf D}\cdot{\bf D}\right|^2 + (A^2+\left|{\bf D}\right|^2)^2 + (\nabla\cdot{\bf A})^2 + \left|(\nabla-i{\bf A})\cdot{\bf D}\right|^2
    \nonumber\\
    & + 4\text{Im}\left[({\bf q}\cdot{\bf D}^*)(\nabla-i{\bf A})\cdot{\bf D}\right] - 4({\bf q}\cdot{\bf A})(A^2+\left|{\bf D}\right|^2) + 4({\bf q}\cdot{\bf A})^2 + 4\left|{\bf q}\cdot{\bf D}\right|^2,
    \nonumber\\
    \left|\partial_i\hat{\psi}\cdot\partial_j\hat{\psi}\right|^2 =&\ \left|D_i D_j\right|^2,
    \nonumber\\
    \text{Re}\left[(\partial_i-iq_i)\hat{\psi}^*\cdot(\partial_j+iq_j)\hat{\psi}\right]^2 =&\ \text{Re}\left[D_i^*D_j + A_iA_j - q_i A_j - q_j A_i + q_i q_j\right]^2,
\end{align}
where we used
\begin{align}
    (\partial_i\hat{\psi})\cdot(\partial_j\hat{\psi}^*) =&\ (\hat{\psi}^*\cdot\partial_i\hat{\psi})(\hat{\psi}\cdot\partial_j\hat{\psi}^*) + (\lh\cdot\partial_i\hat{\psi})(\lh\cdot\partial_j\hat{\psi}^*) = A_i A_j + D_i D_j^*,
    \nonumber\\
    (\partial_i\hat{\psi})\cdot(\partial_j\hat{\psi}) =&\ (\lh\cdot\partial_i\hat{\psi})(\lh\cdot\partial_j\hat{\psi}) = D_i D_j,
    \nonumber\\
    \hat{\psi}^*\cdot(\nabla^2\hat{\psi}) =&\ \partial_i(\hat{\psi}^*\cdot\partial_i\hat{\psi}) - (\nabla\hat{\psi}^*)\cdot(\nabla\hat{\psi}) = -i\nabla\cdot{\bf A} - A^2 - |{\bf D}|^2,
    \nonumber\\
    \hat{\psi}\cdot(\nabla^2\hat{\psi}) =&\ - (\nabla\hat{\psi})^2 = -D^2,
    \nonumber\\
    \lh\cdot(\nabla^2\hat{\psi})=&\ \partial_i(\lh\cdot\partial_i\hat{\psi}) - (\nabla\lh)\cdot(\nabla\hat{\psi}) = (\nabla-i{\bf A})\cdot{\bf D}.
\end{align}
We use this description to formulate the coplanar spin-smectic $\sigma$-model.

\section{Complementary description of \texorpdfstring{$O(N)$}{O(N)} smectic model and the
  \texorpdfstring{$\lambda_3$}{lambda3} term}\label{app:H_Toner}
A complementary description of the spin-smectic was proposed by John
Toner, by starting with
\begin{equation}
\cH_{Toner} = J\left[(\nabla^2 \Sv)^2 
- 2q_0^2(\hat{t}\cdot\nabla\Sv)^2\right] + \oh  K_0(\nabla\hat{t})^2,
\label{H_toner}
\end{equation}
where the nematic-like field $\hat{t}$ ensures underlying rotational
symmetry. Using this model to derive Goldstone mode theory for the
coplanar state (\ref{eq:S_coplanar}), at harmonic level in the small
expansion of the orthonormal triad in terms of three angles
$\chi,\theta_n,\theta_m$
\begin{eqnarray}
\nh&\approx& \eh_1 - \chi\eh_2 + \theta_n\eh_3,\\ 
\mh&\approx& \eh_2 + \chi\eh_1 + \theta_m\eh_3,
\end{eqnarray}
one obtains
\begin{eqnarray}
\cH_{Toner} &=& \oh B (\partial_z\chi)^2 + \oh K(\nabla_\perp^2\chi)^2
+\oh B(\nabla\theta_n)^2+\oh B(\nabla\theta_m)^2.
\end{eqnarray}
That is, the smectic phonon (spiral phase angle $\chi$) is indeed
smectic-like, but the two out-of-plane fluctuations of the triad are
XY-like. This form does not exhibit the out-of-plane instability of
the ``soft'' sigma model (\ref{eq:H_J_coplanar}).

This suggests that the first derivation is from a non-generic model
and therefore misses some important couplings that will appear in a
more general model. Examination of the model proposed by Toner shows
that it contains a new quartic term, that when averaged over $\hat{t}$
gives
\begin{eqnarray}
\langle t_i t_j t_k t_l\rangle
(\partial_i\Sv\cdot\partial_j\Sv)(\partial_k\Sv\cdot\partial_l\Sv)
&\propto&
|\nabla\Sv|^4+ 2(\partial_i\Sv\cdot\partial_j\Sv)^2,
\end{eqnarray}
where we used isotropy of the $\hat{t}$ probability distribution. The
second term is the stabilizing $\lambda_3$ quartic term that enters
crucially for the nonlinear planar spin-smectic $\sigma$-model.

\section{Calculation details of the spin-smectic structure
  factor\label{app:structure_fac}}

In this appendix, we calculate the spin-smectic structure factor in a
Gaussian approximation using the angular representation of the
spin-density waves in Sec.~\ref{sec:ang_rep}. We note that the analysis below neglects effects of nonlinearities that may lead to a crossover to
a nontrivial spin-smectic fixed point, thereby modifying these predictions at long scales. We first calculate
thermal averages of the Goldstone modes,
$\varphi, \varphi' = \chi, \theta, \phi$ (or superposition of them):
\begin{align}
    \langle\cos\varphi\rangle =&\ \oh \langle e^{i\varphi}\rangle + \oh \langle e^{-i\varphi}\rangle = \oh e^{-\oh \langle\varphi^2\rangle} + \oh e^{-\oh \langle\varphi^2\rangle}
    \nonumber\\
    =&\ e^{-\oh \langle\varphi^2\rangle},
    \nonumber\\
    \langle\sin\varphi\rangle =&\ \frac{1}{2i}\langle e^{i\varphi}\rangle - \frac{1}{2i}\langle e^{-i\varphi}\rangle = \frac{1}{2i}e^{-\oh \langle\varphi^2\rangle} -  \frac{1}{2i}e^{-\oh \langle\varphi^2\rangle}
    \nonumber\\
    =&\ 0.
\end{align}
Accordingly the following two-point correlators are given by
\begin{align}
    \langle\cos\varphi({\bf x})\cos\varphi'({\bf x}')\rangle =&\ \oh \langle\cos[\varphi({\bf x})+\varphi'({\bf x}')]\rangle + \oh \langle\cos[\varphi({\bf x})-\varphi'({\bf x}')]\rangle
    \nonumber\\
    =&\ e^{-\oh \langle\varphi^2\rangle-\oh \langle\varphi'^2\rangle}\cosh C_{\varphi\varphi'}({\bf x}-{\bf x}'),
    \nonumber\\
    \langle\sin\varphi({\bf x})\sin\varphi'({\bf x}')\rangle =&\ \oh \langle\cos[\varphi({\bf x})-\varphi'({\bf x}')]\rangle - \oh \langle\cos[\varphi({\bf x})+\varphi'({\bf x}')]\rangle
    \nonumber\\
    =&\ e^{-\oh \langle\varphi^2\rangle-\oh \langle\varphi'^2\rangle}\sinh C_{\varphi\varphi'}({\bf x}-{\bf x}'),
    \nonumber\\
    \langle\cos\varphi({\bf x})\sin\varphi'({\bf x}')\rangle =&\ \oh \langle\sin[\varphi({\bf x})+\varphi'({\bf x}')]\rangle - \oh \langle\sin[\varphi({\bf x})-\varphi'({\bf x}')]\rangle
    \nonumber\\
    =&\ 0,
\end{align}
where 
\begin{align}
    C_{\varphi\varphi'}({\bf x}-{\bf x}') = \langle\varphi({\bf x})\varphi'({\bf x}')\rangle.
\end{align}
In general, all Goldstone modes are coupled and therefore $C_{\varphi\varphi'}\neq 0$ for any $\varphi$ and $\varphi'$. Below, we neglect the coupling between $\theta$, $\phi$, and $\chi$, a valid approximation at low energies. Specifically, we consider the correlators
\begin{align}
    C_{\chi\chi}({\bf x}),\quad C_{\theta\theta}({\bf x}),\quad C_{\phi\phi}({\bf x}),
\end{align}
neglecting all others. The structure factor $\mathcal{S}({\bf q})$ is given by the Fourier transform of the spin-spin correlation function, (\ref{eq:structure_fun_def}).

\subsection{Collinear state}

For the collinear state (\ref{eq:S_collinear}), the spin-spin correlation function in momentum space is given by
\begin{align}
    \int_{{\bf x},{\bf x'}}e^{i{\bf q}\cdot({\bf x}-{\bf x'})}\langle S_\alpha({\bf x})S_\beta({\bf x'})\rangle =&\  \int_{{\bf x},{\bf x'}}e^{i{\bf q}\cdot({\bf x}-{\bf x'})}\langle \hat{n}_\alpha({\bf x})\hat{n}_\beta({\bf x'})\rangle \langle \cos[{\bf q}_0\cdot{\bf x}+\chi({\bf x})]\cos[{\bf q}_0\cdot{\bf x}'+\chi({\bf x}')]\rangle
    \nonumber\\
    = &\ \frac{V}{4}\int_{{\bf x}}\left[e^{i({\bf q}-{\bf q}_0)\cdot{\bf x}}+e^{i({\bf q}+{\bf q}_0)\cdot{\bf x}}\right]D_{\alpha\beta}({\bf x}) e^{-\oh q_0^2C_{sm}({\bf x})},
\end{align}
where in the second line above we dropped the term proportional to $e^{\pm i{\bf q}\cdot({\bf x}+{\bf x}')}$ that will average to zero for ${\bf q}_0\neq 0$ and then renamed ${\bf x}-{\bf x}'\to {\bf x}$. In the above, $C_{sm}({\bf x})$ is defined in (\ref{eq:C_sm}) and $D_{\alpha\beta}({\bf x}) = \langle \hat{n}_\alpha({\bf x})\hat{n}_\beta(0)\rangle$, where by using (\ref{eq:collinear_ang_rep}) the matrix elements are given by
\begin{align}
    D_{11} =&\  e^{-\langle\theta^2\rangle-\langle\phi^2\rangle}\cosh C_{\theta\theta}({\bf x}) \cosh C_{\phi\phi}({\bf x}),
    \nonumber\\
    D_{22} =&\  \oh e^{-\langle\theta^2\rangle-\langle\phi^2\rangle}\cosh C_{\theta\theta}({\bf x}) \sinh C_{\phi\phi}({\bf x}),
    \nonumber\\
    D_{33} =&\ e^{-\langle\theta^2\rangle}\sinh C_{\theta\theta}({\bf x}),
\end{align}
and all others vanish. Keeping terms up to quadratic order in Goldstone
modes, we have
\begin{align}
    D({\bf x}) \approx &\ \left(\begin{array}{ccc}
    1 - \langle\theta^2\rangle - \langle\phi^2\rangle& 0 & 0 \\
    0 & C_{\phi\phi}({\bf x}) & 0 \\
    0 & 0 & C_{\theta\theta}({\bf x})
    \end{array}\right).
\end{align}
This harmonic approximation is consistent with the one we made on the
Hamiltonian and it gives the physically reasonable expression that is
symmetric around the axis $\hat{e}_1$, i.e., $D_{22}=D_{33}$. Within
this Gaussian approximation the structure factor is then given by
\begin{align}\label{eq:structure_fac_collinear}
    \mathcal{S}({\bf q}) =&\  \frac{1}{V}\sum_\alpha\int_{{\bf x},{\bf x}'}e^{i{\bf q}\cdot({\bf x}-{\bf x'})}\langle S_\alpha({\bf x})S_\alpha({\bf x'})\rangle,
    \nonumber\\
    \approx &\ \frac{1}{4}\int_{{\bf x}}\left[e^{i({\bf q}-{\bf q}_0)\cdot{\bf x}}+e^{i({\bf q}+{\bf q}_0)\cdot{\bf x}}\right]e^{-\oh q_0^2C_{sm}({\bf x})-\oh C_{xy,\theta}({\bf x})-\oh C_{xy,\phi}({\bf x})},
\end{align}
where $C_{xy,\theta}({\bf x})$ and $C_{xy,\phi}({\bf x})$ are defined
in (\ref{eq:C_xy}) and in the last line we rewrote small Goldstone
mode fluctuations in an exponential form, $\text{Tr}D({\bf x}) = e^{-\frac{1}{2}C_{xy,\theta}({\bf x})-\frac{1}{2}C_{xy,\phi}({\bf x})}$.

\subsection{Coplanar state}

Repeating above analysis and approximations for the coplanar state
(\ref{eq:S_coplanar}), the spin-spin correlation function in momentum
space is then given by
\begin{align}
    \int_{{\bf x},{\bf x'}}e^{i{\bf q}\cdot({\bf x}-{\bf x'})}\langle S_\alpha({\bf x})S_\beta({\bf x'})\rangle =&\ \frac{1}{4}\int_{{\bf x},{\bf x'}}e^{i{\bf q}\cdot({\bf x}-{\bf x'})}\left(\langle\hat{\psi}_\alpha({\bf x})\hat{\psi}_\beta^*({\bf x'})\rangle e^{i{\bf q}_0\cdot({\bf x}-{\bf x'})} + \langle\hat{\psi}_\alpha({\bf x})\hat{\psi}_\beta({\bf x'})\rangle e^{i{\bf q}_0\cdot({\bf x}+{\bf x'})} + \text{H.c.}\right)
    \nonumber\\
    = &\ \frac{V}{4}\int_{{\bf x}}\left[e^{i({\bf q}+{\bf q}_0)\cdot{\bf x}}D_{\alpha\beta}({\bf x}) + e^{i({\bf q}-{\bf q}_0)\cdot{\bf x}}D^*_{\alpha\beta}({\bf x})\right]e^{-\oh q_0^2C_{sm}({\bf x})},
\end{align}
where
$D_{\alpha\beta}({\bf x}) = \langle\hat{\psi}_\alpha({\bf
  x})\hat{\psi}_\beta^*(0)e^{i[\chi({\bf x})-\chi(0)]}\rangle$ and
H.c. denotes the Hermitian conjugate. Using
(\ref{eq:coplanar_ang_rep}), the matrix elements are given by
\begin{align}
    D_{11} =&\ \oh e^{-\langle\theta^2\rangle-\langle\phi^2\rangle}\sinh C_{\theta\theta}({\bf x}) \cosh C_{\phi\phi}({\bf x}) + \oh e^{-\langle\phi^2\rangle}\sinh C_{\phi\phi}({\bf x}),
    \nonumber\\
    D_{22} =&\ \oh e^{-\langle\theta^2\rangle-\langle\phi^2\rangle}\sinh C_{\theta\theta}({\bf x}) \sinh C_{\phi\phi}({\bf x}) + \oh e^{-\langle\phi^2\rangle}\cosh C_{\phi\phi}({\bf x}),
    \nonumber\\
    D_{33} =&\ \oh e^{-\langle\theta^2\rangle}\cosh C_{\theta\theta}({\bf x}),
    \nonumber\\
    D_{32} =&\ -D_{23} = \frac{i}{2} e^{-\oh \langle\theta^2\rangle-\oh \langle\phi^2\rangle},
\end{align}
with all others contributions vanishing. For small Goldstone mode
fluctuations, we obtain a physically reasonable expression that is
symmetric around the $\hat{e}_1$ axis,
\begin{align}
    D \approx &\ \oh\left(\begin{array}{ccc}
    C_{\theta\theta}({\bf x}) + C_{\phi\phi}({\bf x}) & 0 & 0 \\
    0 & 1 - \langle\phi^2\rangle & -i\left(1 - \oh \langle\theta^2\rangle - \oh \langle\phi^2\rangle\right) \\
    0 & i\left(1 - \oh \langle\theta^2\rangle - \oh \langle\phi^2\rangle\right) & 1 - \langle\theta^2\rangle
    \end{array}\right).
\end{align}
With these approximations the structure factor is then given by
\begin{align}\label{eq:structure_fac_coplanar}
    \mathcal{S}({\bf q}) =&\ \frac{1}{V}\sum_\alpha\int_{{\bf x},{\bf x}'}e^{i{\bf q}\cdot({\bf x}-{\bf x'})}\langle S_\alpha({\bf x})S_\alpha({\bf x'})\rangle
    \nonumber\\
    =&\ \frac{1}{4}\sum_\alpha \int_{{\bf x}}\left[e^{i({\bf q}-{\bf q}_0)\cdot{\bf x}}+e^{i({\bf q}+{\bf q}_0)\cdot{\bf x}}\right]D_{\alpha\alpha}({\bf x}) e^{-\oh q_0^2C_{sm}({\bf x})}
    \nonumber\\
    \approx &\ \frac{1}{4}\int_{{\bf x}}\left[e^{i({\bf q}-{\bf q}_0)\cdot{\bf x}}+e^{i({\bf q}+{\bf q}_0)\cdot{\bf x}}\right] e^{-\oh q_0^2C_{sm}({\bf x})-\frac{1}{4}C_{xy,\theta}({\bf x})-\frac{1}{4}C_{xy,\phi}({\bf x})},
\end{align}
where in the last line we rewrote the small Goldstone mode
fluctuations as exponential form, $\text{Tr}D({\bf x}) = e^{-\frac{1}{4}C_{xy,\theta}({\bf x})-\frac{1}{4}C_{xy,\phi}({\bf x})}$. 

The expressions for the collinear (\ref{eq:structure_fac_collinear}) and coplanar (\ref{eq:structure_fac_coplanar}) structure factors, together with the Goldstone mode correlators (\ref{eq:C_sm}) and (\ref{eq:C_xy}), lead to the same asymptotic form (\ref{eq:structure_fac}), characterized by the following double-power-law peaks
\begin{align}
    P({\bf k}) = P_{sm}({\bf k}) + P_{xy}({\bf k}).
\end{align}
In the above, the leading power-law contribution is from the smectic Goldstone mode fluctuations, given by
\begin{align}
    P_{sm}({\bf k}) =&\ D_0\int_{|{\bf x}|\gg a}e^{i{\bf k}\cdot{\bf x}}e^{-\oh q_0^2C_{sm}({\bf x})}
    \nonumber\\
    \sim &\ D_0\int dx_\parallel d^2x_\perp e^{i{\bf k}\cdot{\bf x}}\times\left\lbrace\begin{array}{cc}
        \frac{1}{x_\perp^{2\eta}},\quad \text{for } x_\perp\gg\sqrt{\lambda |x_\parallel|}, \\
        \frac{1}{x_\parallel^\eta},\quad  \text{for } x_\perp\ll\sqrt{\lambda |x_\parallel|},
    \end{array}\right.
    \nonumber\\
    \sim &\ D_0\left\lbrace\begin{array}{cc}
   \frac{1}{|{\bf k}_\perp|^{4-2\eta}}, & \text{for }k_\parallel = 0,   \\
   \frac{1}{|k_\parallel|^{2-\eta}}, & \text{for }k_\perp = 0,
    \end{array}\right.
\end{align}
where $D_0 = \text{Tr}D(x\to\infty)$ is the Debye-Waller factor, the temperature dependent exponent $\eta$ is given by (\ref{eq:eta}), and we used the change of variable $x_\parallel = x_\perp^2$ to get the last line. On the other hand, the sub-leading power-law contribution is a consequence of the combined effects of the smectic and XY Goldstone mode fluctuations, given by
\begin{align}
    P_{xy}({\bf k}) \sim &\ \int_{|{\bf x}|\gg a}e^{i{\bf k}\cdot{\bf x}}e^{-\oh q_0^2C_{sm}({\bf x})}\left[C_{\theta\theta}({\bf x}) + C_{\phi\phi}({\bf x})\right]
    \nonumber\\
    \sim &\ \frac{T}{\kappa}\int dx_\parallel d^2x_\perp e^{i{\bf k}\cdot{\bf x}}\times\left\lbrace\begin{array}{cc}
        \frac{1}{x_\perp^{2\eta+1}},\quad \text{for } x_\perp\gg\sqrt{\lambda |x_\parallel|}, \\
        \frac{1}{x_\parallel^{\eta+1}},\quad  \text{for } x_\perp\ll\sqrt{\lambda |x_\parallel|},
    \end{array}\right.
    \nonumber\\
    \sim &\ \frac{T}{\kappa}\left\lbrace\begin{array}{cc}
   \frac{1}{|{\bf k}_\perp|^{2(1-\eta)+\frac{\eta}{1+\eta}}}, & \text{for }k_\parallel = 0,   \\
   \frac{1}{|k_\parallel|^{1-\eta+\frac{1}{1+2\eta}}}, & \text{for }k_\perp = 0,
    \end{array}\right.
\end{align}
where for simplicity we chose $\kappa=\kappa_\parallel=\kappa_\perp$ and we used the change of variable $x_\parallel^{\eta+1} = x_\perp^{2\eta+1}$ to get the last line.

\bibliography{ref}

\end{document}